\documentclass[
preprint,
nofootinbib,
 amsmath,amssymb,
 aip,
]{revtex4-2}

\usepackage{graphicx}
\usepackage{dcolumn}
\usepackage{bm}

\usepackage{graphicx}
\usepackage{amsbsy}
\usepackage{latexsym}
\usepackage{xcolor}
\usepackage{graphicx}
\usepackage{psfrag}
\usepackage[normalem]{ulem}
\usepackage{bm}
\usepackage{hyperref}
\usepackage[SF,sf]{subfigure}
\usepackage[nottoc]{tocbibind}
\hypersetup{colorlinks=true, linkcolor=red, citecolor=blue, urlcolor=blue}
\usepackage{float}
\usepackage{multirow}
\usepackage{caption}
\newcommand{\xx}{\mathbf{x}}

\def\ii{\mathbf{i}}
\def\j{\mathbf{j}}
\def\e{\varepsilon}

\begin{document} 

\title{Interactions between droplets in immiscible liquid suspensions and the influence of surfactants}

\author{A.~J.~Archer}
\affiliation{Department of Mathematical Sciences and Interdisciplinary Centre for Mathematical Modelling, Loughborough University, Loughborough LE11 3TU, United Kingdom}

\author{D.~N.~Sibley}
\affiliation{Department of Mathematical Sciences and Interdisciplinary Centre for Mathematical Modelling, Loughborough University, Loughborough LE11 3TU, United Kingdom}

\author{B.~D.~Goddard}
\affiliation{School of Mathematics and Maxwell Institute for Mathematical Sciences, University of Edinburgh, Edinburgh EH9 3FD,
United Kingdom}


\begin{abstract}
We develop a general method for determining the effective interaction potential between two or more droplets suspended within a fluid phase. Our approach is based on classical density functional theory. Here, we apply the method to determine the interaction potential between oil droplets suspended in water and also consider the influence of adding a third species, alcohol. This ternary mixture is that found in the ouzo beverage. The ouzo system exhibits spontaneous emulsification when the neat spirit is mixed with water. The oil emulsion that forms has been observed to be surprisingly long-lived. Here we show that the alcohol in the system does indeed play a role in making the droplets more stable, by decreasing the oil-water interfacial tension and therefore also the strength of the attractive interactions between droplets. Within our theory, the surfactant nature of the alcohol can be enhanced without changing the bulk fluid thermodynamics. In fact, our theory can be used to model surfactant mixtures. In this model, the effective interaction between pairs of oil droplets can become repulsive, with a free-energy barrier to droplets merging, thus making them stable.
\end{abstract}

\date{\today}

\maketitle

\section{Introduction}

The behaviour of liquid droplets suspended within a liquid of another type is important in numerous daily situations.
Of course, for this to occur, the two liquids must be immiscible.
For example, various foods and condiments involve mixtures of water and oils \cite{bai2021recent}.
Salad dressings, containing oil and vinegar that do not mix, typically have to be shaken vigorously before serving, and afterwards the oil droplets generally aggregate again fairly rapidly.
For this reason, an emulsifier is added to mayonnaise, to allow the oil and vinegar to remain mixed and, similarly, many other foods are formulated as an emulsion, with the addition of surfactants to stabilise the system \cite{kralova2009surfactants, ozturk2016progress}.

Mixtures of immiscible liquids are also important in oil recovery, numerous chemical manufacturing processes, washing, waste water treatment and many other instances.
In all these situations, one key factor that determines how the two liquids separate over time \cite{bray2002theory, onuki2002phase}, is the effective interaction between pairs of droplets of the minority liquid phase, through the surrounding bulk of the majority liquid phase.
This is particularly true when in the regime where the phase separation dynamics is dominated by droplets moving together via diffusion and/or hydrodynamics and joining \cite{pototsky2014coarsening, stierle2021hydrodynamic, archer2023stability}, rather than coarsening via Ostwald ripening \cite{lifshitz1961kinetics}.
Generally, the effective interactions between droplets are strongly attractive, which is what drives the aggregation.
However, as we show here, by adding surfactants the effective interaction potentials between droplets can become weaker and, for strong surfactants, the effective interaction potentials can even become repulsive, rendering the droplets stable against aggregation for much longer timescales.

Here, we present a method for determining the effective interaction potential $\Delta\Omega_2(L)$ between pairs liquid droplets suspended in another liquid phase, where $L$ is the distance between the centres of the drops.
The subscript `2' indicates the number of droplets being considered; we also briefly consider the case of three droplets and our approach can in principle be applied to as many as needed.
Our approach for determining $\Delta\Omega_2(L)$ is based on classical density functional theory (DFT) \cite{evans1979nature, hansen2013theory} and is quite general; it can be applied to any mixtures of immiscible liquids, so long as a reliable free energy functional exists for the system of interest.
Here, we apply our approach to the specific case of oil droplets suspended in water.
Hence, throughout we refer to the two liquids as `oil' and `water', but our approach can easily be adapted to other systems.

We also consider the influence of adding a third species of molecules, specifically ethanol, which is a weak surfactant, because it decreases the oil-water surface tension \cite{archer2024experimental, sibley2025coexisting}. We also examine the influence of a strong surfactant species, which has a much greater affinity to the oil-water interfaces than ethanol.
Thus, our DFT is one for ternary mixtures.
The surfactant-like properties of the third component are controlled in our model by adding terms to our DFT that allow us to vary the affinity of the third species to the oil-water interface, without changing the bulk phase behaviour of the ternary mixture.
This enables us to directly assess the influence of having varying amounts of the third species at the oil-water interfaces, decoupling these effects from changes related to the bulk fluid phase behaviour.

The small oil droplets that we have in mind can become dispersed within the water when oil-water mixtures are vigorously stirred or shaken (e.g.\ in salad dressing).
They can also spontaneously form via the so-called ouzo effect, when water is added to a stable mixture of oil, water and alcohol \cite{Vitale2003, solans2016spontaneous}.
The alcohol in ouzo enables the small amount of anise oil to remain mixed with the water.
But when further water is added, this leads to the system entering the unstable (demixing) portion of the phase diagram, which can lead to droplets of the minority oil phase forming \cite{archer2024experimental}.
These then subsequently coalesce over time, either via aggregation or via Ostwald ripening.

In the particular case of ouzo, the droplets of oil that form when water is added to the neat spirit mixture can be stable over surprisingly long time-scales \cite{sitnikova2005spontaneously, archer2024experimental}.
This observation in part motivates the present study, where one of our goals here is to assess whether the alcohol is a strong enough surfactant to stabilise the oil droplets, leading to the observed long life-times and stability.
We use a version of the DFT developed in \cite{archer2024experimental} for the ouzo mixture, to determine the effective interactions between oil droplets.
We find that this DFT does predict a decrease in the strength of the effective interaction potential between the oil droplets, since the added alcohol leads to a decrease in the oil-water interfacial tension, and also a density enhancement of the alcohol at the oil-water interface.
However, the additional alcohol in the system and, in particular, the molecules adsorbed at the interface are not sufficient to lead to any repulsive barriers in $\Delta\Omega_2(L)$, which would stabilise the droplets.

As mentioned above, we also introduce (somewhat ad-hoc) additional terms into the free energy that do not change the bulk phase behaviour of the mixture, but do allow us to vary of the affinity between the `alcohol' and the oil-water interface.
We find that as the affinity is increased, the `alcohol' adsorbed at the interface becomes increasingly surfactant-like in character, and is indeed able to stabilise the oil droplets, by leading to the appearance of a repulsive barrier in $\Delta\Omega_2(L)$.
The addition of surfactants leading to increased stability has been observed in experiments on emulsions formed via the ouzo effect, where they are seen to remain stable for longer periods \cite{chen2025impact}.
Whilst the simple model presented here is plausible, further work is required to determine whether in reality alcohol adsorption at the oil water interface leads to the stabilization of anise oil droplets in water.

The approach presented here builds upon previous work using DFT to determine the effective interaction between pairs of colloidal particles suspended in a liquid or between a single colloid and some other object, such as the container wall \cite{stark2004capillary, andrienko2004capillary, archer2005solvent, hopkins2009solvent, okamoto2013attractive, chacko2017solvent}.
In these works, the colloids are treated as external potentials, with the density distribution of the surrounding liquid calculated via DFT.
The two colloids are fixed, with the centres a distance $L$ apart. DFT is used to determine the density profiles of the surrounding fluid and also the thermodynamic grand potential of the whole system, $\Omega$ \cite{evans1979nature, hansen2013theory}.
This calculation is then repeated for a range of values of $L$, yielding the effective solvent mediated potential between the pair of colloids as  \cite{archer2005solvent, hopkins2009solvent}:
\begin{equation}
    \Delta\Omega_2(L)\equiv\Omega(L)-\Omega(L\to\infty).
    \label{eq:1}
\end{equation}
The central idea here is to perform the same kind of calculation, but replacing the external potentials due to the colloids with weak, slowly varying Gaussian potentials. The range of the Gaussian potentials is chosen so that they are negligible outside of the oil droplets.
Moreover, we choose these potentials to act solely on the oil.
The external potentials for the water and alcohol/surfactant are zero everywhere.
The centres of these two Gaussian potentials are set to be a distance $L$ apart.
The amplitude is chosen to be fairly small: strong enough to hold the oil droplets in place, but weak enough to hardly change the density distribution of the oil droplets.
In particular, we make sure that the Gaussian external potentials do not distort in any way the shape of the oil-water interfaces.
Thus, there are some parallels between the present work and that in Refs.~\onlinecite{archer2005solvent, archer2003solvent, archer2002microscopic}, where the focus was on determining the solvent mediated potential between pairs of large Gaussian particles, interacting strongly with a surrounding liquid (also of Gaussian particles). The Gaussian potentials of Refs.~\onlinecite{archer2005solvent, archer2003solvent, archer2002microscopic} interact with the surrounding fluid much more strongly than the very weak potentials considered here.

The effective potentials $\Delta\Omega_2(L)$ that we calculate  have two different branches.
Where these branches cross (each branch with a different gradient) it results in a jump in $f_2=-\partial\Delta\Omega_2(L)/\partial L$, corresponding to a jump in the force as droplets merge.
Qualitatively very similar results have been observed in atomic force microscope (AFM) experiments measuring the force between droplets \cite{israelachvili2011intermolecular}.
In such experiments, one of the droplets is attached to the AFM cantilever, while the other droplet is attached to a surface.
The DFT that we use here is based on a very simplified coarse-grain description of the molecular interactions, making direct comparison with experiment impossible.
For example, we neglect to consider the effect of any charged species or charge screening that might be present in the system.
The excellent book \cite{israelachvili2011intermolecular} gives a good discussion of how such factors can influence the intermolecular forces between droplet interfaces.
Nonetheless, putting these caveats aside, the results of our theory are in line with AFM measurements of the force between between oil droplets in water \cite{gunning2004atomic}, which show that the force jumps at contact and is strongly attractive as the drops merge.
AFM measurements have also been made for water droplets surrounded by oil (toluene) \cite{sun2021unraveling}.
This study also examines the influence of surfactants adsorbed at the oil-water interface.
They find repulsive forces between the drops when they are covered with surfactant.
Molecular dynamics simulations have also been used to calculated the potential of mean force, obtaining results qualitatively very similar to those that we present here for our surfactant model \cite{sun2021unraveling}.

Other background relevant molecular dynamics simulation studies include Ref.~\onlinecite{arbabi2025collision}, which focusses on the collision dynamics of surfactant-laden droplets, comparing with the collisions of pure water droplets.
They observe bridging between droplets, with configurations very reminiscent of some that we present below.
In another computer simulation study \cite{pak2018free}, the authors obtain the free energy as a function of the distance between the centres of mass of a pair of water droplets.
Again, these are similar to the potentials $\Delta\Omega_2(L)$ obtained here.
Another recent computer simulation study \cite{inada2025molecular} uses dissipative particle dynamics for oil droplets with various different surfactants on the surfaces of the drops. They find pair-interaction forces $f_2$ between the droplets that are qualitatively very similar to what we find here, with a jump in the force at contact.

In the final part of this paper we use dynamical density functional theory (DDFT) \cite{marconi1999dynamic, archer2004dynamical, espanol2009derivation, te2020classical} together with our DFT for ternary oil-alcohol-water mixtures to study the coarsening dynamics following a quench of the uniform mixture into an unstable part of the phase diagram (i.e.\ inside the spinodal).
We consider cases where the dynamics involves the formation and subsequent merging of droplets of the minority oil-rich phase, surrounded by a background of the water-rich majority phase.
This situation allows to observe the effect on the coarsening dynamics of varying the affinity of the `alcohol' towards the oil-water interfaces (i.e.\ we vary the surfactant-like properties of the `alcohol').
We see that the more strongly it adsorbs at the oil-water interfaces, the more slowly the coarsening dynamics proceeds.
This alternate way of assessing the stability of the oil droplets tallies nicely with the understanding obtained from our investigations of the effective potential between droplets $\Delta\Omega_2(L)$.

Previous studies that use (D)DFT and related theories that we should mention include Ref.~\onlinecite{liu2025anomalous}, where results for polymers dissolved in various solvents are presented.
These are used as a coarse-grained model for biomolecular condensates in intracellular environments.
For the cases where the solvent is a poor solvent (i.e. immiscible with the polymer), these exhibit potentials between the polymer droplets that are akin to those that we find here.
Hydrodynamic DDFT has also been used for droplet coalescence \cite{stierle2021hydrodynamic}, showing how the density and fluid velocity vary during the merger of nitrogen, propane and other hydrocarbon droplets.
In Ref.~\onlinecite{heinen2022droplet}, molecular dynamics and phase field modelling (which may be viewed as a form of DDFT) are compared, for the coalescence of pairs of argon droplets.
The authors report excellent agreement over the whole coalescence process.

This paper is structured as follows: In Sec.~\ref{sec:2}, we give an overview of the thermodynamics of droplet interactions. Then, in Sec.~\ref{sec:3} we outline briefly our DFT for ternary oil-water-alcohol mixtures, with Sec.~\ref{subsec:surfactant} describing the terms we add to the free energy in order to model surfactants.
In Sec.~\ref{sec:4} we discuss the bulk fluid phase behaviour and present the phase diagram.
In Sec.~\ref{sec:results} we present our results for the interaction potentials between oil droplets, $\Delta\Omega_2(L)$.
In Sec.~\ref{subsec:pure_oil_water} we discuss the pure oil-water system, then in Sec.~\ref{subsec:alcohol} we discuss the influence of alcohol on $\Delta\Omega_2(L)$, before  presenting results for $\Delta\Omega_2(L)$ for our surfactant model.
In Sec.~\ref{sec:3_drops} we present results for the three-droplet effective interaction potential $\Delta\Omega_3$.
In Sec.~\ref{sec:dynamics} we show DDFT results for the dynamics following a quench into the spinodal region of the phase diagram.
Finally, in Sec.~\ref{sec:conc}, we make a few concluding remarks.

\section{Thermodynamics of droplet interactions}
\label{sec:2}

The thermodynamics of finite sized droplets of one liquid species suspended in another fluid phase is best analysed in the semi-grand canonical ensemble, with the following discussion following a similar line of argument to that made in Ref.~\onlinecite{hopkins2009solvent}.
Thus, we treat the majority species within the droplets (the oil) in the canonical ensemble, fixing the total number of molecules in the system.
In contrast, we treat the bulk liquid majority species (the water) and any third species, such as alcohol or surfactant, grand canonically, fixing the chemical potentials of these two species.
The Landau (grand) potential of the system without any droplets is
\begin{equation}
    \Omega_0=-p_{\rm w}V,
\end{equation}
where $p_{\rm w}$ is the pressure of the bulk liquid (water rich) phase and $V$ is the volume of the system.
Here, we use subscripts `w' and `o' to denote respectively the water and oil coexisting bulk phases\footnote{Strictly speaking, we should refer to these as the water-rich and oil-rich coexisting phases, since of course for entropic reasons there is always a little of the other species dissolved in each phase.}.
When there is one droplet in the system, then the grand potential is the following sum of volume- and surface-related contributions
\begin{eqnarray}
    \Omega_1&=& -p_{\rm w}\left(V-\frac43\pi R^3\right)
    -p_{\rm o}\frac43\pi R^3
    +4\pi R^2\gamma(R)\nonumber \\
    &=& \Omega_0+\frac43\pi R^3(p_{\rm w}-p_{\rm o})
    +4\pi R^2\gamma(R),
\end{eqnarray}
where $R$ is the radius of the (spherical) droplet, $p_{\rm o}$ is the pressure of the liquid inside the droplet (the oil rich phase) and $\gamma$ is the interfacial tension for the oil-water interface. The surface tension of the spherical droplet can be written as
\begin{equation}
    \gamma(R)=\gamma(\infty)\left(1-\frac{2\delta}{R}+\cdots\right),
    \label{eq:toleman}
\end{equation}
where $\gamma(\infty)\equiv\gamma_{\rm ow}$ is the interfacial tension for the planar oil-water interface and $\delta$ is the Tolman length, which is of order the size of the molecules \cite{rowlinson1982molecular}.
Thus, the approximation $\gamma(R)\approx\gamma_{\rm ow}$ for all $R$ becomes increasingly good as $R$ becomes larger.
Typically, such droplets arise when the system is at or near to liquid-liquid phase coexistence, i.e.\ when $p_{\rm w}\approx p_{\rm o}$. Thus, in this limit, the free energy for the droplet being in the system is
\begin{equation}
    \Omega_1\approx\Omega_0 + 4\pi R^2\gamma_{\rm ow}.
    \label{eq:5}
\end{equation}
In other words, the energy to insert a single oil droplet of radius $R$ into the system $(\Omega_1-\Omega_0)$ is largely determined by the size of the drop and the value of the surface tension, $\gamma_\mathrm{ow}$.

Similarly, one can consider the case when there are two droplets in the system, separated by a distance $L$.
When the two droplets are far apart from each other, then the insertion free energy is just double that for inserting a single droplet, $2(\Omega_1-\Omega_0)$.
This result of course assumes that both droplets are of equal size, with radii $R$.
The grand potential of the system is then just
\begin{eqnarray}
    \Omega_2(L\to\infty)&=& -p_{\rm w}\left(V-2\frac43\pi R^3\right)
    -2p_{\rm o}\frac43\pi R^3
    +8\pi R^2\gamma(R)\nonumber\\
    &=&2\Omega_1-\Omega_0.
    \label{eq:Omega_2}
\end{eqnarray}
As the droplets approach one another, i.e.\ as the centre-to-centre distance $L$ is decreased, then Eq.~(\ref{eq:Omega_2}) is no longer a good approximation.
When $L \approx 2R$ one should expect that (unless there are strong surfactants present in the system) the droplets merge and become a single droplet.
The effective interaction potential between a pair of droplets may be defined as [c.f.\ Eq.~\eqref{eq:1}]
\begin{equation}
\label{eq:8}
    \Delta\Omega_{2}(L)\equiv\Omega_2(L)-2\Omega_1+\Omega_0.
\end{equation}
We should emphasise this is of course also a function of the radii of the two droplets.
Our approach could be used to describe droplets of different radii, but we don't consider that case here.

One can also generalise the above to determine the effective interaction potential between multiple droplets.
For example, the effective three-body potential between three droplets is
\begin{equation}
    \Delta\Omega_{3}(\xx_1,\xx_2,\xx_3)\equiv\Omega_3(\xx_1,\xx_2,\xx_3)-3\Omega_1+2\Omega_0,
    \label{eq:3body}
\end{equation}
where $\Omega_3(\xx_1,\xx_2,\xx_3)$ is the grand potential of the system with three droplets centred  at points $\xx_1$, $\xx_2$ and $\xx_3$.
In Sec~\ref{sec:3_drops} we display examples of $\Delta\Omega_{3}$, for specific droplet configurations.

As explained in the Introduction, the approach we take here is to calculate $\Delta\Omega_{2}(L)$ and $\Delta\Omega_3$ using DFT via a constrained minimization approach.
The constraint consists of applying a small external potential that fixes the centres of the droplets at specified distances apart.
We choose the potentials to act solely on the oil phase, thus for a pair of oil droplets, we set the external potential acting on the oil phase to be
\begin{equation}
    \Phi_{\rm o}(\xx)=-A\exp\left(-\frac{(x-L/2)^2+y^2+z^2}{w^2}\right)-A\exp\left(-\frac{(x+L/2)^2+y^2+z^2}{w^2}\right),
    \label{eq:ext_pot_o}
\end{equation}
where $\xx=(x,y,z)$ and where $A>0$ is the amplitude and $w$ is the range of the potentials.
For the case of three droplets, we generalise the above potential to include a third Gaussian.
The other two potentials acting on the water `w' and the alcohol `a', are set to zero,
\begin{equation}
    \Phi_{\rm w}(\xx)=\Phi_{\rm a}(\xx)=0.
    \label{eq:ext_pot_w}
\end{equation}
We set the range $w$ of the Gaussian potentials \eqref{eq:ext_pot_o} to be a little less than the radius $R$ of the droplets.
In this paper we treat the liquid via lattice-DFT (not continuum DFT) and so we replace the potentials \eqref{eq:ext_pot_o} and \eqref{eq:ext_pot_w} by their discrete lattice equivalents.
However, before we describe this, in the following section we first briefly describe the simple lattice-DFT that we use.

\section{DFT model}
\label{sec:3}

The DFT we use is based on that developed recently in Refs.~\onlinecite{archer2024experimental, sibley2025coexisting} for the ternary oil-water-alcohol (ouzo) system.
The free energy is constructed by assuming the system can be mapped onto a discrete lattice.
This built on earlier work for one- and two-component systems \cite{kierlik2001capillary, woo2001mean, gouyet2003description, woywod2003phase, robbins2011modelling, monson2012understanding,
schneider2014filling, hughes2014introduction, hughes2015liquid, chacko2015two, chalmers2017dynamical, kikkinides2022connecting, archer2023stability, areshi2024binding}.
The essence of the approach is to consider the liquid mixture to be within a space that is discretised onto a three-dimensional cubic lattice, with lattice spacing $\sigma$.
We assume that the volumes at each lattice site are of roughly the size of one of the molecules and are just the right size that each cube can contain the centre of mass of no more than one molecule at any moment in time.
We denote the position of each lattice site via the index $\ii =(i,j,k)$, where $i$, $j$ and $k$ are integers.
The ensemble average densities of each of the three species $\{\rm a, o, w\}=\{$alcohol, oil, water$\}$ at lattice site $\ii$ are then denoted $n_\ii^{\rm a}$, $n_\ii^{\rm o}$ and $n_\ii^{\rm w}$, respectively. These probabilities satisfy the constraints $0<n_\ii^p<1$ for all $p\in \{\rm a,o,w\}$ and $(n_\ii^{\rm a}+n_\ii^{\rm o}+n_\ii^{\rm w})<1$. The second condition, that the sum of the probabilities for site $\ii$ to be occupied is less than 1, comes from the constraint that at most one molecule of either a,o,w can be at that site at any given moment.
The Helmholtz free energy can then be approximated as \cite{archer2024experimental}
\begin{align}
F  = & k_B T \sum_\ii \Big[
        n^{\rm a}_\ii \log n^{\rm a}_\ii
    + n^{\rm o}_\ii \log n^{\rm o}_\ii
    + n^{\rm w}_\ii \log n^{\rm w}_\ii
    \nonumber \\
    & \qquad \qquad + (1 - n^{\rm a}_\ii - n^{\rm o}_\ii-n^{\rm w}_\ii) \log (1 - n^{\rm a}_\ii - n^{\rm o}_\ii-n^{\rm w}_\ii)\Big]
  \nonumber\\
  &- \sum_{\ii,\j}\Big(
\frac{1}{2} \e_{\ii\j}^{\rm aa} n_{\ii}^{\rm a} n_{\j}^{\rm a}
+\frac{1}{2} \e_{\ii\j}^{\rm oo} n_{\ii}^{\rm o} n_{\j}^{\rm o}
+\frac{1}{2} \e_{\ii\j}^{\rm ww} n_{\ii}^{\rm w} n_{\j}^{\rm w}\nonumber\\
&\qquad \qquad
+ \e_{\ii\j}^{\rm wa} n_\ii^{\rm w} n_\j^{\rm a}
+ \e_{\ii\j}^{\rm wo} n_\ii^{\rm w} n_\j^{\rm o}
+ \e_{\ii\j}^{\rm ao} n_\ii^{\rm a} n_\j^{\rm o}
\Big)
\nonumber
\\
&\quad
+ \sum_\ii \left(\Phi_\ii^{\rm a} n_\ii^{\rm a} + \Phi_\ii^{\rm o} n_\ii^{\rm o}+ \Phi_\ii^{\rm w} n_\ii^{\rm w}\right),
\label{eq:helmholtz}
\end{align}
where $k_B$ is the Boltzmann constant and $T$ is the temperature. The first four terms in Eq.~\eqref{eq:helmholtz} (those involving the logarithms) are entropic in origin: recall that the Helmholtz free energy $F=-TS+U$, where $S$ is the entropy and $U$ is the internal energy, so of course the remaining terms in Eq.~\eqref{eq:helmholtz} are energetic in origin.
The term in the second line acts as a constraint enforcing that the total density $n^{\rm a}_\ii + n^{\rm o}_\ii + n^{\rm w}_\ii < 1$. It originates from the core repulsions between the particles, taking that particular form due to the particle-`hole' symmetry of particles constrained to be on a lattice\cite{hughes2014introduction}.
The terms in the last line of Eq.~\eqref{eq:helmholtz} are those due to any external potentials $\Phi_\ii^p$ acting on the three different species.
In the work here, only the potential acting on the oil is non-zero, being used to constrain the oil droplets to be a distance $L$ apart.
It is the lattice generalization of Eqs.~\eqref{eq:ext_pot_o} and \eqref{eq:ext_pot_w}, given by
\begin{equation}
    \Phi_\ii^p=\Phi_p(\xx=\ii\sigma).
\end{equation}

The six matrices $\e_{\ii\j}^{pq}$, with values $\e_{\ii\j}^{pq} = \epsilon_{pq} c_{\mathbf{ij}}$ and where $\{p,q\}\in \{\rm a,o,w\}$, correspond to the discrete (on the lattice) pair interaction potentials between particles at different lattice sites \cite{archer2024experimental}.
These terms all have the form
\begin{equation}
  - \sum_{\ii,\j}\e_{\ii\j}^{pq} n_\ii^p n_\j^q  = - \epsilon_{pq} \sum_{\ii,\j}c_{\ii\j} n_\ii^p n_\j^q,
\end{equation}
with an additional prefactor of $1/2$ when $p\neq q$.
Note the minus signs, so that $\e_{\ii\j}^{pq}>0$ corresponds to an attractive pair interaction. The overall strength of each of the potentials is determined by the parameters $\epsilon_{pq}$, for $\{p,q\}\in \{\rm a,o,w\}$.
Here, we follow Refs.~\onlinecite{chalmers2017modelling, chalmers2017dynamical, archer2024experimental} and choose the tensor
\begin{equation}
\label{eq:c_ij}
c_{\mathbf{ij}}  = 
  \begin{cases} 
   1 & \text{if }\mathbf{j}\in {NN \mathbf{i}}, \\
    \frac{3}{10}    & \text{if }\mathbf{j}\in {NNN \mathbf{i}}, \\
    \frac{1}{20}    & \text{if }\mathbf{j}\in {NNNN \mathbf{i}}, \\
        0   & \text{otherwise},
         \end{cases}
\end{equation}
where $NN\mathbf{i}$, $NNN\mathbf{i}$ and $NNNN\mathbf{i}$ denote the nearest neighbours of $\mathbf{i}$, next nearest neighbours of $\mathbf{i}$ and next-next nearest neighbours of $\mathbf{i}$, respectively.

The specific choice in Eq.~\eqref{eq:c_ij} is made so that liquid-liquid interfaces and the corresponding density profiles hardly depend on the orientation with respect to the underlying lattice \cite{archer2023stability, areshi2024binding}.
So, as we see below, the cross-section of the oil droplets suspended in water are close to being circular, as they should be!
This choice to have the discretised pair potential to be as isotropic as possible turns out to be equivalent to requiring that the discretisation of the Laplace operator introduces as few lattice-discretisation artefacts as possible \cite{robbins2011modelling, kumar2004isotropic, chalmers2017modelling}.
This is because the interaction terms can also be written as
\begin{equation}
 -\sum_{\ii,\j}c_{\ii\j} n_\ii^p n_\j^q = - \epsilon_{pq}\frac{12\sigma^2}{5} \sum_{\ii} n_\ii^p \nabla^2n_\ii^p
  -10\epsilon_{pq}\sum_\ii n_\ii^p n_\j^q,
 \label{eq:lattice_to_continuum}
\end{equation}
where $\nabla$ should be understood as a finite difference approximation for the gradient operator, with step size equal to the lattice spacing $\sigma=1$. For a uniform bulk fluid with constant densities, the first term on the right hand side of Eq.~\eqref{eq:lattice_to_continuum} is zero, while the second term reduces to the usual bulk mean-field approximation. See the Appendix for more details of the derivation of the result in Eq.~\eqref{eq:lattice_to_continuum}.

Completing the mapping of the discrete system onto the continuum (see the Appendix for details), by replacing $\ii\sigma\to\xx$, $\sigma=1$, $\sum_\ii\to\int d\xx$ and $n^p_\ii\to n_p(\xx)$, we obtain the following expression for the pair interaction terms:
\begin{eqnarray}
  - \sum_{\ii,\j}\e_{\ii\j}^{pq} n_\ii^p n_\j^q  \approx \int \left[ \frac{12}{5}\epsilon_{pq} \nabla n_p(\xx)\cdot \nabla n_q(\xx)-10\epsilon_{pq} n_p(\xx) n_q(\xx)\right]d\xx.
  \label{eq:lattice_to_continuum_2}
\end{eqnarray}
Note that the prefactor 10 in the last term is obtained from $\sum_\j c_{\ii\j}=10$ -- see Eq.~\eqref{eq:c_ij}.
Note also that the species label `$p$' on $n_p(\mathbf{x})$ has moved from superscript to subscript, as we go from the lattice to the continuum.
Using Eq.~\eqref{eq:lattice_to_continuum_2}, we can map the Helmholtz free energy of the system \eqref{eq:helmholtz} to the following functional
\begin{align}
F  = \int \bigg[ &f(n_{\rm a},n_{\rm o},n_{\rm w})
+\frac{12}{5}\bigg(
\frac{1}{2}\epsilon_{\rm aa} (\nabla n_{\rm a})^2
+\frac{1}{2}\epsilon_{\rm oo} (\nabla n_{\rm o})^2
+\frac{1}{2}\epsilon_{\rm ww} (\nabla n_{\rm w})^2\nonumber \\
&+ \epsilon_{\rm wo} (\nabla n_{\rm w})\cdot (\nabla n_{\rm o})
+ \epsilon_{\rm wa} (\nabla n_{\rm w})\cdot (\nabla n_{\rm a})
+ \epsilon_{\rm oa} (\nabla n_{\rm o})\cdot (\nabla n_{\rm a})
\bigg)\nonumber \\
&+ \Phi_{\rm a}n_{\rm a} + \Phi_{\rm o} n_{\rm o}+ \Phi_{\rm w} n_{\rm w}\bigg]d\xx,
\label{eq:helmholtz_continuum}
\end{align}
where the bulk free energy term is given by
\begin{align}
\label{eq:f_bulk}
f &=  k_BT\big[n_{\rm a} \log{n_{\rm a}}+n_{\rm o} \log{n_{\rm o}}+n_{\rm w} \log{n_{\rm w}}\nonumber\\
& \qquad \qquad + (1 - n_{\rm a} - n_{\rm o} - n_{\rm w}) \log(1 - n_{\rm a} - n_{\rm o} - n_{\rm w})\big]
\nonumber
\\ 
&\qquad - 5\epsilon_{\rm aa}(n_{\rm a})^2
- 5\epsilon_{\rm oo}(n_{\rm o})^2
- 5\epsilon_{\rm ww}(n_{\rm w})^2\nonumber\\
&\qquad - 10\epsilon_{\rm wa} n_{\rm w} n_{\rm a}
- 10\epsilon_{\rm wo} n_{\rm w} n_{\rm o}
- 10\epsilon_{\rm ao} n_{\rm a} n_{\rm o}.
\end{align}
The above continuum free energy functional \eqref{eq:helmholtz_continuum} is what is typically referred to as a `square-gradient approximation' for ternary mixtures \cite{evans1979nature, hansen2013theory}; see also Ref.~\onlinecite{schilling2002wetting} for another way to write this.
We must emphasise that the lattice free energy \eqref{eq:helmholtz} and a discretisation of the continuum Eq.~\eqref{eq:helmholtz_continuum} are identical, as long as the lattice spacing for the discretisation is $\sigma=1$.

\subsection{Strong surfactant modelling}
\label{subsec:surfactant}

As discussed in the introduction, we also add terms to the free energy in order to change the character of the alcohol to make it more akin to a stronger surfactant. We must emphasise that these additional terms do not change the bulk fluid phase behaviour and only change the surface tension and other interfacial behaviour. This is done by adding the following pair of terms to the free energy
\begin{equation}
 F_{s}=   -\epsilon_{3{\rm o}}\int n_{\rm a}(\nabla n_{\rm o})^2d\xx
    -\epsilon_{3{\rm w}}\int n_{\rm a}(\nabla n_{\rm w})^2d\xx,
    \label{eq:F_s}
\end{equation}
where of course 
$\nabla n_p = \left(\frac{\partial n_p}{\partial x},\frac{\partial n_p}{\partial y},\frac{\partial n_p}{\partial z}\right)$, so that the free energy now becomes $F+F_{s}$.
The central idea in choosing the form of $F_s$ is to add terms that lower the free energy if the density of the alcohol $n_{\rm a}$ is higher at the oil-water interface, i.e.\ where there are gradients in the density profiles of the oil and the water.
The coefficients $\epsilon_{3{\rm o}}$ and $\epsilon_{3{\rm w}}$ control the overall strength of these terms, with the subscript `3' to remind us that these terms are cubic in the densities.
To evaluate the partial derivatives in Eq.~\eqref{eq:F_s} on the lattice, we use the following expression for the partial derivative in the $x$-direction,
\begin{eqnarray}
    \frac{\partial n_p}{\partial x}&=&\frac{1}{10\sigma}\bigg(
    \left[n^p_{(i+1,j,k)}-n^p_{(i-1,j,k)}\right]\nonumber\\
   & &\qquad\qquad +\left[n^p_{(i+1,j+1,k)}-n^p_{(i-1,j+1,k)}\right]
    +\left[n^p_{(i+1,j-1,k)}-n^p_{(i-1,j-1,k)}\right]\nonumber\\
   & &\qquad\qquad+\left[n^p_{(i+1,j,k+1)}-n^p_{(i-1,j,k+1)}\right]
    +\left[n^p_{(i+1,j,k-1)}-n^p_{(i-1,j,k-1)}\right]
    \bigg),
    \label{eq:dn_dx}
\end{eqnarray}
with corresponding expressions for the other two partial derivatives, in the $y$ and $z$ directions. Note that if one were to replace the expression in Eq.~\eqref{eq:dn_dx}, with a much simpler one, such as the central difference expression $\frac{\partial n_p}{\partial x}=\frac{1}{2\sigma}\left[n^p_{(i+1,j,k)}-n^p_{(i-1,j,k)}\right]$, then we find that this leads to the interfacial tension having a strong dependence on the orientation with respect to the underlying lattice -- the droplets become cube-like -- which, of course, is undesirable.
This approach/attitude is akin to that used previously in Refs.~\onlinecite{robbins2011modelling, kumar2004isotropic, chalmers2017modelling} to obtain Eq.~\eqref{eq:c_ij}.
With the approximation in Eq.~\eqref{eq:dn_dx}, the droplets are almost perfectly circular in cross-section and so this is the expression we use throughout here when modelling surfactants.

\section{Bulk phase behaviour}
\label{sec:4}

In Refs.~\onlinecite{archer2024experimental, sibley2025coexisting}, it was demonstrated that the free energy in Eq.~\eqref{eq:helmholtz} describes well the bulk phase behaviour and surface tension of the ouzo ternary mixture.
This was achieved by making appropriate choices for the six pair interaction parameters $\epsilon_{pq}$, for $\{p,q\}\in\{{\rm a,o,w}\}$.
Here, we use similar values for these six parameters, but not exactly the same values.
The reason for changing the values is that the DFT calculations become easier if the system is at a state point where the overall compressibility of the system is a little larger. Or, to put it another way, the DFT is less `stiff' for state points where the probability of finding any given lattice site to be vacant is at least a few percent.
For the alternative $\epsilon_{pq}$ parameters chosen here, a typical value for the total density is $(n_{\rm a}+n_{\rm o}+n_{\rm w})\approx0.97$, i.e.\ is further below 1 than it is for the set of $\beta\epsilon_{pq}$ used in Refs.~\onlinecite{archer2024experimental, sibley2025coexisting}, where $\beta=(k_BT)^{-1}$.
As discussed in Ref.~\onlinecite{archer2024experimental}, the ouzo system can be considered to be essentially incompressible. However, for present purposes, this assumption makes the DFT calculations harder than they need to be. 
Thus, the values of the pair interaction parameters used here are:
\begin{eqnarray}
    \beta\epsilon_{\rm ww} &=& 0.72,\hspace{1cm}
    \beta\epsilon_{\rm wo} = 0.36,\hspace{1cm}
    \beta\epsilon_{\rm oo} = 0.72,\nonumber\\
    \beta\epsilon_{\rm aa} &=& 0.60,\hspace{1cm}
    \beta\epsilon_{\rm aw} = 0.66,\hspace{1cm}
    \beta\epsilon_{\rm oa} = 0.48.
    \label{eq:epsilons}
\end{eqnarray}
Roughly speaking, the above values are obtained by decreasing $\beta\epsilon_{\rm ww}$ by a third compared to the value used in Refs.~\onlinecite{archer2024experimental, sibley2025coexisting}, and then selecting the values of the other five to best match the experimentally observed ouzo phase diagram.
The physics of the system dictates what values should be selected:
As discussed further in Ref.~\onlinecite{archer2024experimental}, the value for $\epsilon_{\rm oo}$ should be roughly the same as that of $\epsilon_{\rm ww}$, but the cross interaction $\epsilon_{\rm wo}$ should be much less, since oil and water do not mix. The values of the alcohol related parameters are dictated by the facts that (i) the alcohol-alcohol intermolecular bonding is weaker than that between water molecules (fewer hydrogen bonds) and (ii) it is observed that the alcohol prefers to be in the water-rich phase over being in the oil-rich phase, hence $\epsilon_{\rm aw}>\epsilon_{\rm oa}$.

\begin{figure}
    \centering
    \includegraphics[width=0.49\textwidth]{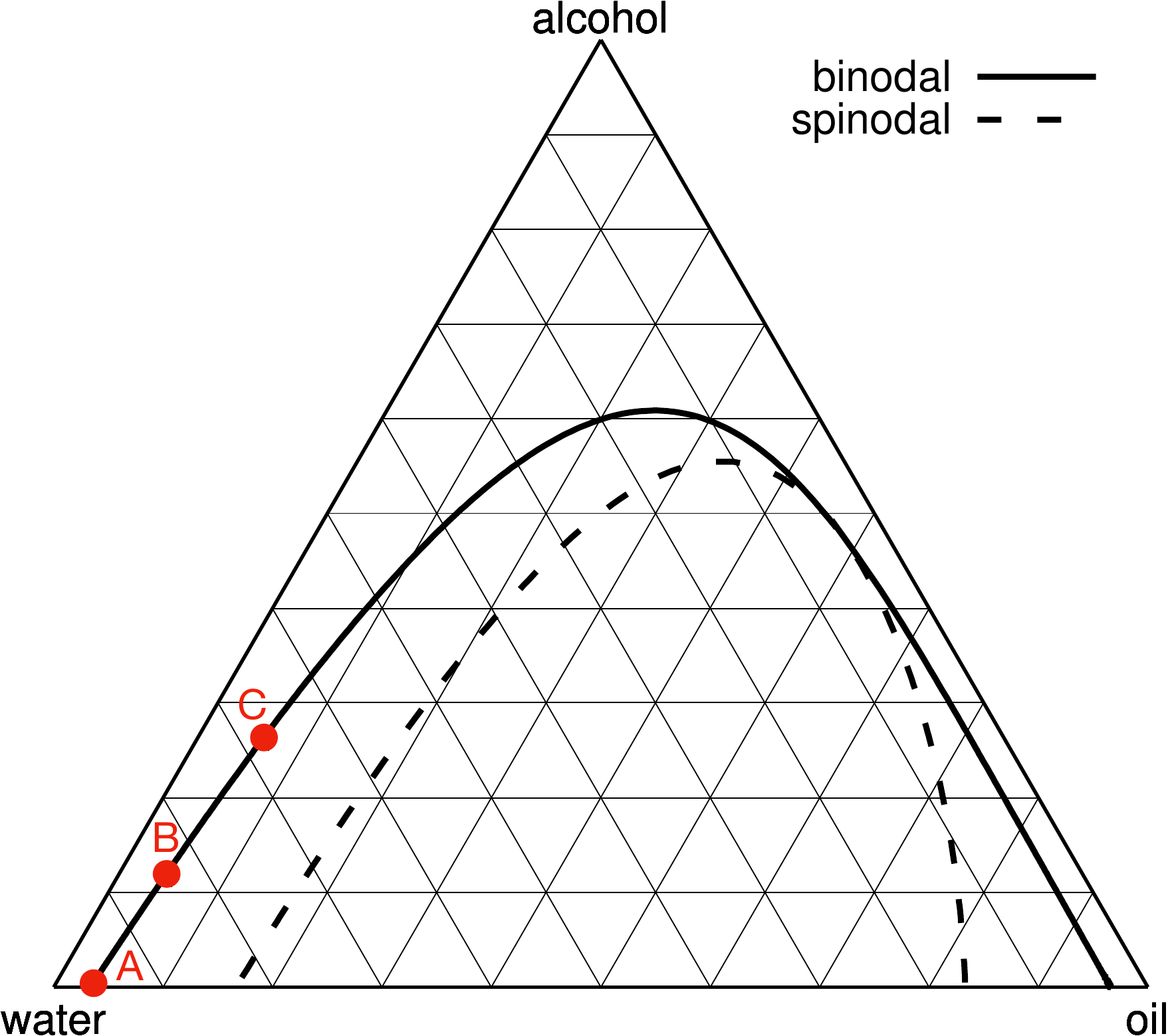}
    \caption{The bulk phase diagram of the ternary oil-water-alcohol (ouzo) system, using the pair interaction parameters given in Eq.~\eqref{eq:epsilons}. Each of the corners correspond to the respective (as labelled) pure liquids, with the concentration of each species decreasing with distance from each respective corner. Below the binodal, the system exhibits two-phase coexistence. Note that in this representation the tie-lines between coexisting state points on the binodals are not horizontal \cite{archer2024experimental}. The critical point is located at the unique point where the binodal and spinodal curves meet tangentially. Note that this phase diagram is for the incompressible mixture, where we assume the total number density $(n_{\rm a}+n_{\rm o}+n_{\rm w})=1$. However, the phase diagram hardly changes if re-calculated for fixed oil chemical potential $\mu_{\rm o}$, in the range $-3.5\lesssim\beta\mu_{\rm o}\lesssim1$. The points A--C correspond to bulk state points where results in Figs.~\ref{fig:pot1}--\ref{fig:three_body} are obtained.}
    \label{fig:phase_diag}
\end{figure}

The bulk liquid phase diagram is calculated using the approach described in Ref.~\onlinecite{archer2024experimental} and is displayed in Fig.~\ref{fig:phase_diag}.
We calculate the binodal, which corresponds to the locus of coexisting phases, and also the spinodal, below which the mixture becomes spontaneously unstable to demixing.
Note also that the phase diagram displayed in Fig.~\ref{fig:phase_diag} is for the incompressible mixture, where we assume $(n_{\rm a}+n_{\rm o}+n_{\rm w})=1$.
One can instead calculate the phase diagram for a fixed value of the oil chemical potential $\mu_{\rm o}$, but when doing this, we find that the phase diagram hardly changes from that displayed in Fig.~\ref{fig:phase_diag}, as long as $\mu_{\rm o}$ is in the rather broad range $-3.5\lesssim\beta\mu_{\rm o}\lesssim1$.
Note also that we use the following standard coordinate transform to map the individual densities onto a triangular ternary phase diagram:
\begin{align}
x=\frac{1}{2}\frac{2n_{\rm o}+n_{\rm a}}{n_{\rm a}+n_{\rm o}+n_{\rm w}}, \nonumber \\
y=\frac{\sqrt{3}}{2}\frac{n_{\rm a}}{n_{\rm a}+n_{\rm o}+n_{\rm w}}.
\label{eq:coord_transform}
\end{align}
Of course, for the incompressible system, the two denominators in the fractions above are equal to 1.
The main difference between the phase diagram displayed in Fig.~\ref{fig:phase_diag} and the one in Ref.~\onlinecite{archer2024experimental}, is that the binodal curves do not approach as close to the edges of the diagram as they do in Ref.~\onlinecite{archer2024experimental}. In other words, the present DFT predicts that the coexisting oil and water phases have a little more of the other species dissolved within them, than they do in reality.

\section{Results: interaction potential between oil droplets}
\label{sec:results}

DFT is generally formulated as a grand canonical theory \cite{evans1979nature, hansen2013theory} and in this case, for a ternary mixture, the equilibrium fluid density profiles are obtained by minimizing the grand potential functional
\begin{equation}
    \Omega = F -\mu_{\rm a}\int n_{\rm a}d\xx
    -\mu_{\rm o}\int n_{\rm o}d\xx
    -\mu_{\rm w}\int n_{\rm w}d\xx,
    \label{eq:omega}
\end{equation}
where the Helmholtz free energy functional $F$ is given in Eq.~\eqref{eq:helmholtz_continuum} and the chemical potentials of the three species, $\mu_{\rm a}$, $\mu_{\rm o}$ and $\mu_{\rm w}$, respectively, are specified before hand.
However, to obtain stable droplets of the oil within the liquid, one must instead treat the oil phase canonically, fixing the total number of oil molecules in the system to be a predetermined value.
This issue is discussed further in Refs.~\onlinecite{hughes2014introduction, hughes2015liquid} in the context of using DFT to calculate the density profile of stable droplets on planar surfaces.
Note that this semi-grand canonical treatment is needed because the Gaussian potential \eqref{eq:ext_pot_o} is not strong enough to create oil droplets of the desired size in a grand canonical calculation. It is only strong enough to keep their centres of mass fixed in place. If we treated the oil grand canonically, the oil droplets would shrink significantly and in some cases even disappear completely into the reservoir.
The water and alcohol are still treated grand-canonically, i.e.\ by fixing the chemical potentials of these two species.
Thus, we determine the density profiles of the three different species for the case of one or more oil droplets surrounded by the bulk water phase by minimising the following semi-grand free energy
\begin{equation}
    \Omega = F -\mu_{\rm a}\int n_{\rm a}d\xx
    -\mu_{\rm w}\int n_{\rm w}d\xx,
    \label{eq:omega_semi}
\end{equation}
subject to the additional constraint that the total number of oil molecules in the system,
\begin{equation}
    N_{\rm o} = \int n_{\rm o}d\xx,
    \label{eq:N_oil}
\end{equation}
is fixed. Of course, this is mathematically the same as minimising Eq.~\eqref{eq:omega}, but with the Lagrange multiplier $\mu_{\rm o}$ not specified a priori.

An additional point to mention here is that to make our computations easier we treat the system as varying in only two of the Cartesian directions and assume it to be invariant in the third direction, making our computations two dimensional (2D).
Thus, we effectively calculate the potential per unit length between two liquid cylinders, rather than between two spherical droplets.
This is satisfactory for present purposes, because much of the physics revealed for this 2D system qualitatively applies also to 3D droplets.
However, some of the results discussed in Sec.~\ref{sec:2} must be adapted to the 2D situation at hand. Specifically, the 2D analogue of Eq.~\eqref{eq:5}, the excess grand potential for having one droplet (cylinder of oil) in the system, is
\begin{equation}
    \Omega_1-\Omega_0\approx 2\pi R \ell \gamma_{\rm ow}.
    \label{eq:5_in2D}
\end{equation}
where $\ell$ is the length of the cylindrical droplet and $R$ is the radius.
Note that we have assumed $\ell$ is large and so have neglected any contributions from the ends of the cylinders, or (equivalently) assumed there are periodic boundary conditions in the direction parallel to the axes of the cylinders.
For the effective interaction between a pair of droplets (cylinders of oil), Eq.~\eqref{eq:8} still applies, but the 2D equivalent of Eq.~\eqref{eq:Omega_2} is
\begin{equation}
    \Omega_2(L\to\infty)= -p_{\rm w}\left(V-2\pi R^2\ell\right)
    -2p_{\rm o}\pi R^2\ell
    +4\pi R\ell\gamma(R),
    \label{eq:Omega_2_2D}
\end{equation}
where the first two terms involve the volume $\pi R^2\ell$ of the two cylinders and the last term involves the surface area $2\pi R\ell$, neglecting the contribution from the ends.
The 2D analogue of the external potential \eqref{eq:ext_pot_o} that we use to fix the locations of the centres of the droplets is
\begin{equation}
    \Phi_{\rm o}(\xx)=-A\exp\left(-\frac{(x-L/2)^2+y^2}{w^2}\right)-A\exp\left(-\frac{(x+L/2)^2+y^2}{w^2}\right),
    \label{eq:ext_pot_o_2D}
\end{equation}
which is identical to Eq.~\eqref{eq:ext_pot_o} when $z=0$.
In everything that follows, we set the length $\ell=\sigma=1$, so that when we discuss the effective interaction potential $\Delta\Omega_2(L)$, strictly speaking we are really discussing the effective potential per unit length, $\Delta\Omega_2(L)/\ell$.

\subsection{Pure oil-water system}
\label{subsec:pure_oil_water}

\begin{figure}
    \centering
    \includegraphics[width=0.9\textwidth]{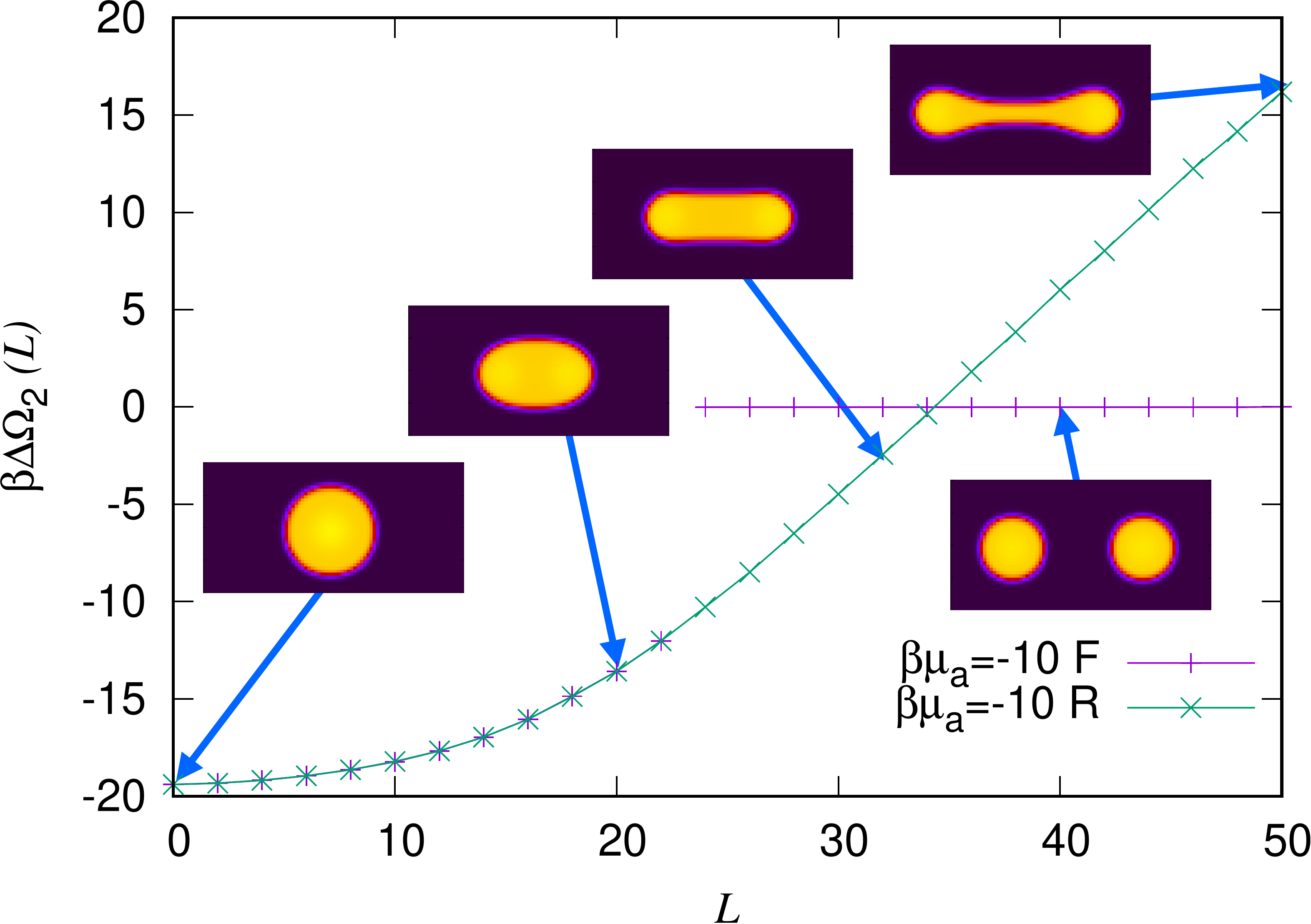}
    \caption{The effective interaction potential $\Delta\Omega_2(L)$ between a pair of oil droplets of diameter $2R\approx20\sigma$ plotted as a function of the distance between the droplet centres $L$, for $\beta\mu_{\rm a}=-10$ (i.e.\ effectively no alcohol in the system) and $\beta\mu_{\rm w}=-3.5$ (the bulk water-rich phase surrounding the droplets corresponds to point A in Fig.~\ref{fig:phase_diag}). The potential $\Delta\Omega_2(L)$ has two branches, one corresponding to the droplets advancing forward `F' towards each other and the other corresponding to a single droplet of diameter $28\sigma$ being pulled apart into two droplets and reversing `R' away from each other. Examples of five typical configurations are indicated (these do not show the whole computational domain). In all cases, the total number of oil molecules in the system of area $80\sigma\times80\sigma$ is fixed to be $N_{\rm o}=800$.}
    \label{fig:pot1}
\end{figure}

We begin by presenting results for the case when the chemical potential of the alcohol $\beta\mu_{\rm a}=-10$, which corresponds to the case where there is essentially no alcohol in the system.
Owing to the fact that this value of $\mu_{\rm a}$ is so low, our model in fact predicts more oil is dissolved in the bulk water phase than alcohol, having number fractions $n_{\rm o}\approx4\times10^{-2}$ and $n_{\rm a}\approx8\times10^{-4}$, respectively.
In Fig.~\ref{fig:pot1} we display the potential $\Delta\Omega_2(L)$, calculated via Eq.~\eqref{eq:8}, between two oil droplets each of diameter $d=2R\approx20\sigma$.
These calculations are performed in a square domain of size $80\sigma\times80\sigma$, with periodic boundary conditions in all directions.
We fix the total number of oil molecules in the system [see Eq.~\eqref{eq:N_oil}] to be 800, while the chemical potential of the water is fixed at $\beta\mu_{\rm w}=-3.5$, a value close to that of bulk liquid-vapour phase coexistence.
It is our choice of $N_{\rm o}$ which determines the value of the droplet diameters $d$, which are subsequently measured from the density profiles.

In Fig.~\ref{fig:pot1}, beginning on the purple solution branch at $L=50$, we see the potential $\Delta\Omega_2(L)\approx0$, corresponding to a pair of droplets that barely influence each other, and $\Delta\Omega_2(L)$ remains very close to zero as $L$ is decreased down to $L=24\sigma$ (at the break in the purple branch).
For $L<24\sigma$, the two droplets come into contact and join to form a single droplet.
Recall that $L$ is both the distance between the centres of the droplets and is also the parameter in the external potential $\Phi_\ii^{\rm o}$ given in Eq.~\eqref{eq:ext_pot_o_2D}, which constrains the centres to be a distance $L$ apart.
The other parameters in the potential $\Phi_\ii^{\rm o}$ are chosen to be $\beta A=0.5$ and $w=5\sigma$, i.e.\ corresponding to a fairly small amplitude and a range that is small enough compared to the radius of the droplets so as to hardly influence the oil-water interfaces.
In other words, our results are insensitive to the precise values of $A$ and $w$.

For $L<24\sigma$, the two droplets join and there is just a single droplet in the system; this is the green solution branch in Fig.~\ref{fig:pot1}, which also has the remainder of the purple branch behind it.
This branch goes right down to $L=0$, i.e.\ the `centres' of the `two' droplets coincide, which of course is just another way of saying there is one droplet.
Turning around on this branch and increasing $L$, which corresponds to pulling apart the single droplet in order to form a pair of droplets, we find first a dumbbell shaped droplet, with a steadily increasing length bridge between the two ends that then breaks for $L>50$, where the system breaks into two droplets, falling back down onto the purple solution branch, corresponding to two separate droplets.

In Fig.~\ref{fig:pot1}, the range over which hysteresis in $\Delta\Omega_2(L)$ occurs is is rather large, $24\leq L\leq50$.
Physically what this corresponds to is a discontinuous jump in the force between the pair of droplets.
We should also expect thermal fluctuations to somewhat round off these discontinuities.
However, given the energy scale for the hysteresis is $\gg k_BT$, the energy scale for thermal fluctuations, we should still expect these jumps to be observable in experiments and to be even more pronounced for larger droplets \cite{archer2005solvent, hopkins2009solvent}.

The minimum of the potential $\Delta\Omega_2(L)$ in Fig.~\ref{fig:pot1} at $L=0$ has the value $\beta\Delta\Omega_2(L=0)=-19.4$.
This corresponds to the free energy difference between there being two isolated small droplets in the system, or being joined to form a single large one.
This difference can also be estimated using Eq.~\eqref{eq:5_in2D}, to give
\begin{align}
    \Delta\Omega_2(L=0) &\approx 
(\Omega_1(d_b)-\Omega_0)-2(\Omega_1(d_s)-\Omega_0)\nonumber\\
&=\pi(d_b-2d_s)\ell\gamma_{\rm ow},
    \label{eq:estimate}
\end{align}
where $d_s$ and $d_b$ are the diameters of the two small droplets and the single big one, respectively.
We determine the diameters $d_s$ and $d_b$ from inspecting the density profiles, defining $d$ as the distance between the oil-water interfaces, measured through the centre of the droplets, and identifying the position of the interfaces to be located on the boundary between pairs of neighbouring lattice sites where one has $n_{\rm o}>0.5$ and the other has $n_{\rm o}<0.5$.
Determined in this manner, the pair of small droplets for $L=50$ have diameter $d_s=20\sigma$ and the single big droplet for $L=0$ has diameter $d_b=28\sigma$.
The remaining quantity required for our estimate of $\beta\Delta\Omega_2(L=0)$ using Eq.~\eqref{eq:estimate} is the oil-water interfacial tension, $\gamma_{\rm ow}$.
This is calculated in the standard way \cite{hughes2014introduction, hansen2013theory}, by calculating the density profiles for the planar oil-water interface at bulk phase coexistence and then from these determining $\gamma_{\rm ow}$ as the excess free energy due to the interface.
For $\beta\mu_{\rm a}=-10$, we obtain the value $\gamma_{\rm ow}=0.479k_BT/\sigma^2$.
Inserting all these values into Eq.~\eqref{eq:estimate}, we obtain the following estimate for the minimum value of the potential
$\beta\Delta\Omega_2(L=0) \approx  -18.1$, which is in good agreement with the value of $\beta\Delta\Omega_2(L=0)=-19.4$ calculated via DFT.
This indicates the validity of Eq.~\eqref{eq:5_in2D} as a rather accurate approximation, even for the relatively small droplets considered here.
This also shows that the volume (pressure) correction terms in Eq.~\eqref{eq:Omega_2_2D} are small and can arguably be neglected.

\subsection{Influence of the alcohol -- weak surfactant}
\label{subsec:alcohol}

\begin{figure}
    \centering
    \includegraphics[width=0.9\textwidth]{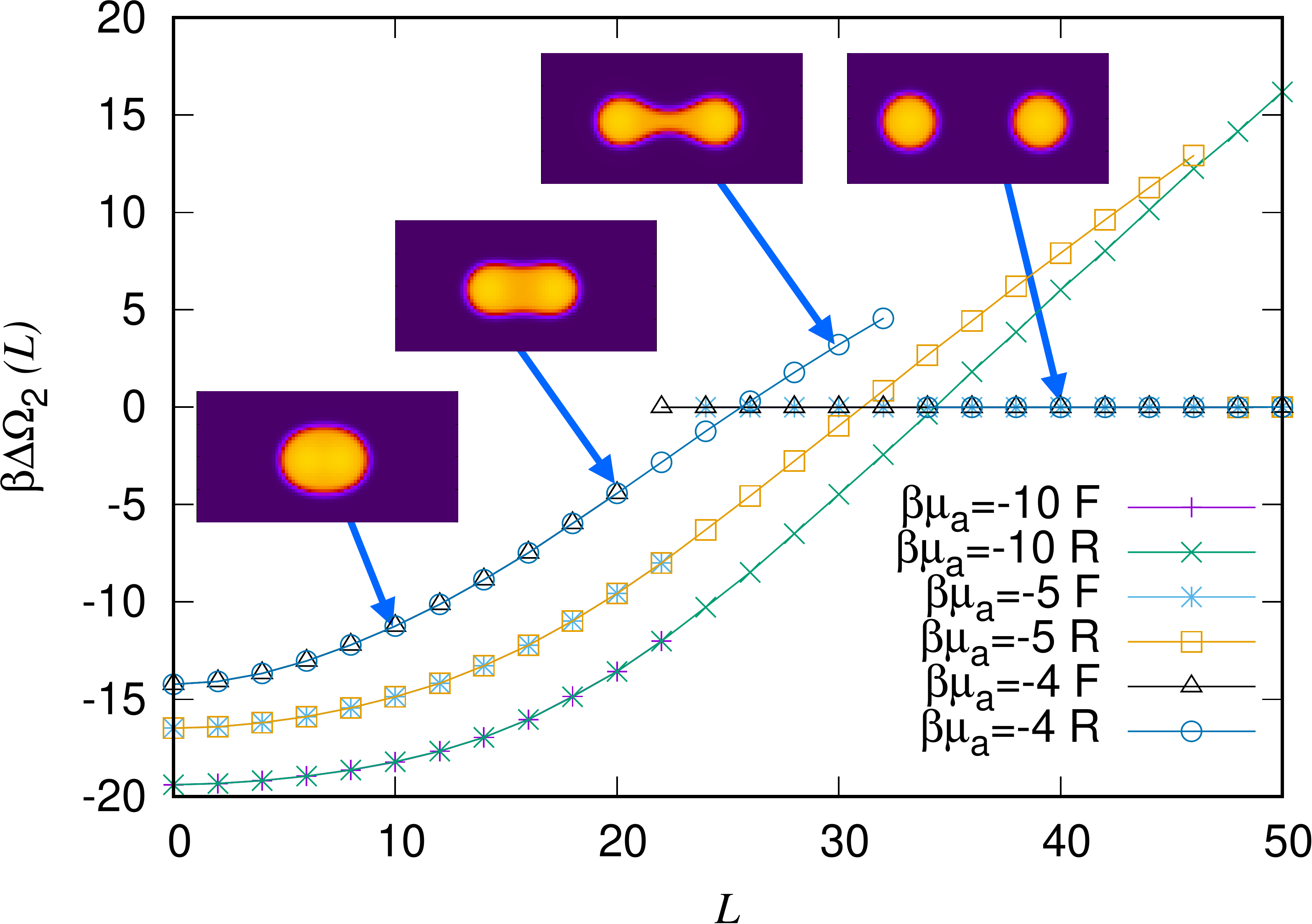}
    \caption{The effective interaction potential $\Delta\Omega_2(L)$ between pairs of oil droplets, plotted as a function of the distance between the droplet centres $L$. These are calculated for the three alcohol chemical potential $\mu_{\rm a}$ values given in the key. The corresponding bulk water-rich phases surrounding the droplets are indicated as points A--C in Fig.~\ref{fig:phase_diag}. The chemical potential of the water $\beta\mu_{\rm w}=-3.5$. The total number of oil molecules $N_{\rm o}=800$ is fixed, in a domain of size $80\sigma\times80\sigma$. The potential $\Delta\Omega_2(L)$ has two branches, one corresponding to the droplets advancing forward `F' towards each other and the other corresponding to the droplets being pulled apart and reversing `R' away from each other. Examples of four typical configurations are indicated.}
    \label{fig:pot2}
\end{figure}

We now discuss results for increasing values of the alcohol chemical potential $\mu_{\rm a}$, i.e.\ for increasing amounts of alcohol in the system.
Figure~\ref{fig:pot2} displays the interaction potential between droplets $\Delta\Omega_2(L)$ for fixed $\beta\mu_{\rm w}=-3.5$ and three different values, $\beta\mu_{\rm a}=-10$, $-5$ and $-4$.
The first of these, $\beta\mu_{\rm a}=-10$, is the value used in Fig.~\ref{fig:pot1} and is repeated in Fig.~\ref{fig:pot2} in order to compare with the results for the other two values of $\mu_{\rm a}$.
The bulk density (number fraction) of the alcohol in the bulk water-rich phase for these three chemical potential values is $n_{\rm a}=8\times10^{-4}$, $0.11$ and $0.25$, respectively.

We see that increasing the amount of alcohol in the system leads to a decrease in both the range and overall strength (i.e.\ depth of the minimum at $L=0$) of $\Delta\Omega_2(L)$.
This is due to two factors: (i) the increased amount of alcohol in the bulk water-rich phase leads to a larger fraction of the oil being dissolved there too.
Thus, the oil droplets become a little smaller as a small fraction of the oil is transferred from the droplets to the bulk.
A consequence of this drop size decrease is a decrease in the range of $\Delta\Omega_2(L)$.
(ii) The extra alcohol in the system leads to a decrease in the surface tension $\gamma_{\rm ow}$, i.e.\ the alcohol is a weak surfactant.
For $\beta\mu_{\rm a}=-10$, as mentioned above, we find the surface tension $\gamma_{\rm ow}=0.479k_BT/\sigma^2$.
Increasing the amount of alcohol, for $\beta\mu_{\rm a}=-5$ we find $\gamma_{\rm ow}=0.395k_BT/\sigma^2$, and for $\beta\mu_{\rm a}=-4$ we obtain $\gamma_{\rm ow}=0.293k_BT/\sigma^2$.

\begin{figure}
    \centering
    \includegraphics[width=0.99\textwidth]{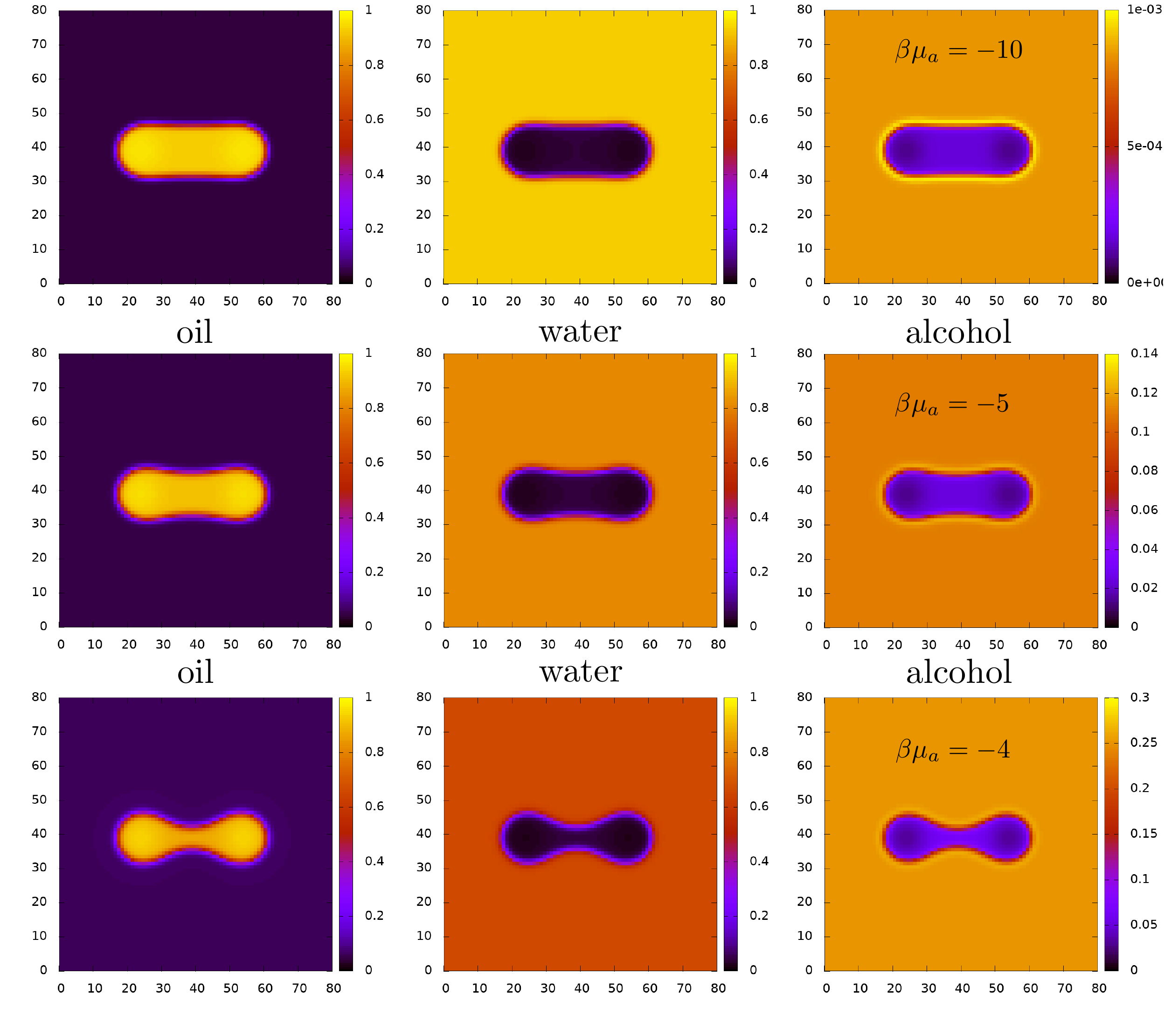} 
    \caption{Density profiles for a fixed amount of oil $N_{\rm o}=800$ and $L=30\sigma$, corresponding to the plots of $\Delta\Omega_2(L)$ displayed in Fig.~\ref{fig:pot2}. These profiles all correspond to the solution branch where the pair of droplets are joined. The chemical potential of the water is $\beta\mu_{\rm w}=-3.5$, while that of the oil increases in each row from top to bottom, as indicated (corresponding to points A--C in Fig.~\ref{fig:phase_diag}, respectively). In each row, the left hand profile is that of the oil, the middle that of the water and the right that of the alcohol.}
    \label{fig:rhos}
\end{figure}

For $\beta\mu_{\rm a}=-5$, from the DFT we find the minimum value of the potential to be $\beta\Delta\Omega_2(L=0)=-16.5$, with the diameter of the pair of small droplets for $L=50\sigma$ being $d_s=18\sigma$, while for $L=0$ the single droplet has diameter $d_b=26\sigma$.
Plugging these values into Eq.~\eqref{eq:estimate} we obtain $\beta\Delta\Omega_2(L=0)\approx-12.4$, which compares reasonably well with the DFT result.
Similarly, for $\beta\mu_{\rm a}=-4$, corresponding to even more alcohol in the system, we obtain $d_s=16\sigma$ and $d_b=24\sigma$, so from Eq.~\eqref{eq:estimate} we obtain $\beta\Delta\Omega_2(L=0)\approx-7.4$, which when compared with the DFT result, $\beta\Delta\Omega_2(L=0)=-14.2$, shows that as the diameter of the droplets decreases, the estimate \eqref{eq:estimate} starts to fare less well.
This is not particularly surprising in view of Eq.~\eqref{eq:toleman}.
Moreover, small droplets have a higher (Laplace) pressure difference between the pressure within and the bulk pressure of the surrounding fluid, so one should expect the pressure terms that are neglected in \eqref{eq:estimate} to be increasingly important for very small droplets.
In contrast, for larger droplets we can be confident that the estimate in Eq.~\eqref{eq:estimate}, and also its 3D analogue, will be increasingly accurate as the droplet radii $R$ increase.

In Fig.~\ref{fig:rhos} we display a selection of density profiles for fixed $N_{\rm o}=800$, $L=30\sigma$ and $\beta\mu_{\rm w}=-3.5$, and for varying $\mu_{\rm a}$, corresponding to the potentials $\Delta\Omega_2(L)$ displayed in Fig.~\ref{fig:pot2}.
These profiles all correspond to the solution branch where the pair of droplets are joined.
The chemical potential of the alcohol increases from top to bottom.
For the case in the top row, the bridge joining the oil droplets has the same diameter as the droplets, but as $\mu_{\rm a}$ is increased, the bridge narrows because the increased alcohol in the system enables a greater amount of the oil to become dissolved in the bulk water-rich phase.
This can also be seen from the phase diagram in Fig.~\ref{fig:phase_diag} and from the changes in the value of the background density in the left hand plots of Fig.~\ref{fig:rhos}.

Another interesting feature of Fig.~\ref{fig:rhos}, which is particularly visible in the right hand alcohol density plots, is the enhancement in the amount of alcohol at the oil-water interface.
Note the changing density (heatmap) colourbar scale.
We have already noted that increasing the amount of alcohol in the system decreases the surface tension of the oil water interface (see also Ref.~\onlinecite{archer2024experimental}), and here we also see a noticeable increase in the density of the alcohol at the oil-water interface.
In view of this enhancement, the fact that the alcohol behaves as a weak surfactant is perhaps not surprising.
That said, in all three cases, the density of alcohol right at the interface is never more than 25\% above the corresponding bulk value, $n_{\rm a}=8\times10^{-4}$, $0.11$ and $0.25$, respectively.

\subsection{Strong surfactant model}
\label{subsec:surfactant_pots}

\begin{figure}
    \centering
    \includegraphics[width=0.99\textwidth]{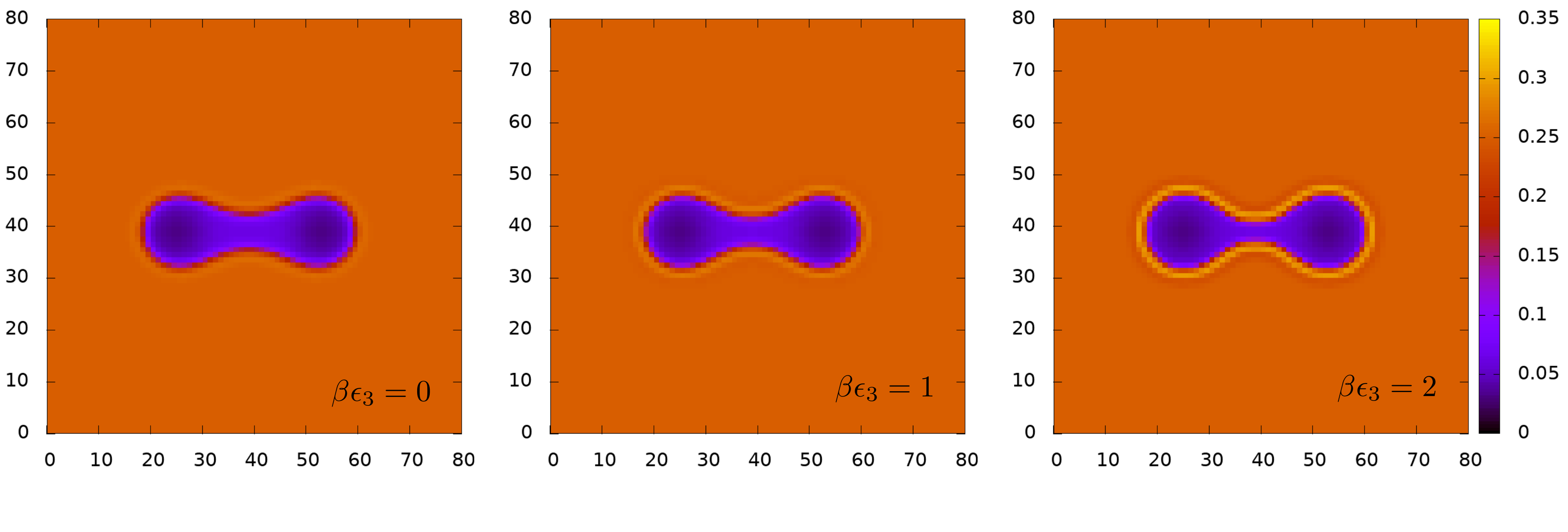}
    
    \caption{Density profiles of the alcohol/surfactant, for varying $\epsilon_3$ and for $N_{\rm o}=800$ and $L=30\sigma$.}
    \label{fig:rho_surfactant}
\end{figure}

\begin{figure}
    \centering
    \includegraphics[width=0.99\textwidth]{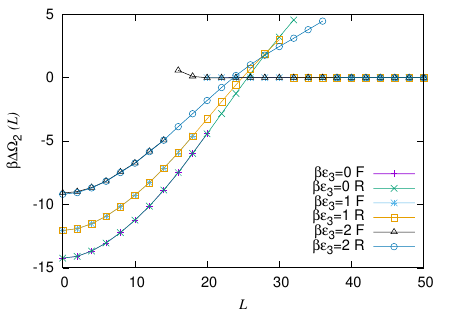}
    
    \caption{Effective interaction potentials between pairs of oil droplets for varying $\epsilon_3$ and for fixed $N_{\rm o}=800$ and $\beta\mu_{\rm a}=-4$ (bulk corresponding to point C in Fig.~\ref{fig:phase_diag}). Note that for $\beta\epsilon_3=2$, the non-bridged branch of the potential between the droplets is repulsive as the drops come close to contact, at $L\approx16$. In other words, for drops to merge, there is a free-energetic barrier to be surmounted.}
    \label{fig:drop_pots_surf}
\end{figure}

Having seen in the previous subsection that the alcohol behaves as a weak surfactant, we now present results for our strong surfactant model, i.e.\ with the free energy terms in Eq.~\eqref{eq:F_s} being non-zero.
The strength of the two terms in Eq.~\eqref{eq:F_s} are controlled by the two parameters $\epsilon_{3{\rm o}}$ and $\epsilon_{3{\rm w}}$.
To simplify, here we set these to be equal, $\epsilon_{3{\rm o}}=\epsilon_{3{\rm w}} \equiv \epsilon_3$.

In Fig.~\ref{fig:rho_surfactant} we display the density profile of the alcohol/surfactant for varying $\epsilon_3$ and fixed $L=30\sigma$, $N_{\rm o}=800$, $\beta\mu_{\rm w}=-3.5$ and $\beta\mu_{\rm a}=-4$. 
Note that the left hand panel of Fig.~\ref{fig:rho_surfactant} is actually the same as the profile displayed in the bottom right of Fig.~\ref{fig:rhos}, but here the heatmap colourbar scale is slightly different. 
In Fig.~\ref{fig:rho_surfactant} all three plots share the same scale bar, so the increase in density at the interface with increasing $\epsilon_3$ is clearly visible.
However, we also see that (as expected) the additional contributions to the free energy do not in any way change the bulk uniform fluid densities.

In Fig.~\ref{fig:drop_pots_surf} we display results for the effective interaction potential $\Delta\Omega_2(L)$ between pairs of oil droplets surrounded by the bulk water-rich phase.
We display three cases: the potential for $\beta\epsilon_3=0$ (also displayed in Fig.~\ref{fig:pot2}), the potential for $\beta\epsilon_3=1$ and also for $\beta\epsilon_3=2$.
The ranges of the three potentials are similar, because the volume of oil in the droplets does not change as $\epsilon_3$ is varied.
In contrast, the depth of the potential, i.e.\ the value of $\Delta\Omega_2(L=0)$, does change significantly, increasing as $\epsilon_3$ is increased.
This is because as $\epsilon_3$ is increased, the oil-water interfacial tension decreases, and so from Eq.~\eqref{eq:estimate} the depth of the potential must become less, with the (negative) minimum value increasing.
However, the most striking aspect to be observed from Fig.~\ref{fig:drop_pots_surf} is that the effective interaction potential $\Delta\Omega_2(L)$ for $\beta\epsilon_3=2$ is actually repulsive.
We see that for $L\geq20\sigma$ the potential for two separate droplets (the forward `F' branch in Fig.~\ref{fig:drop_pots_surf}) is roughly zero. However, as $L$ is decreased down to the value $L=16\sigma$, we see that the free energy increases -- i.e.\ the potential is repulsive.
In other words, the surfactant has stabilized the oil droplets and for them to join a force must be applied to push them together to overcome the free energy barrier due to the adsorbed surfactant layers at the interfaces.

\section{Three-droplet interactions}
\label{sec:3_drops}

\begin{figure}
    \centering
    \includegraphics[width=0.99\textwidth]{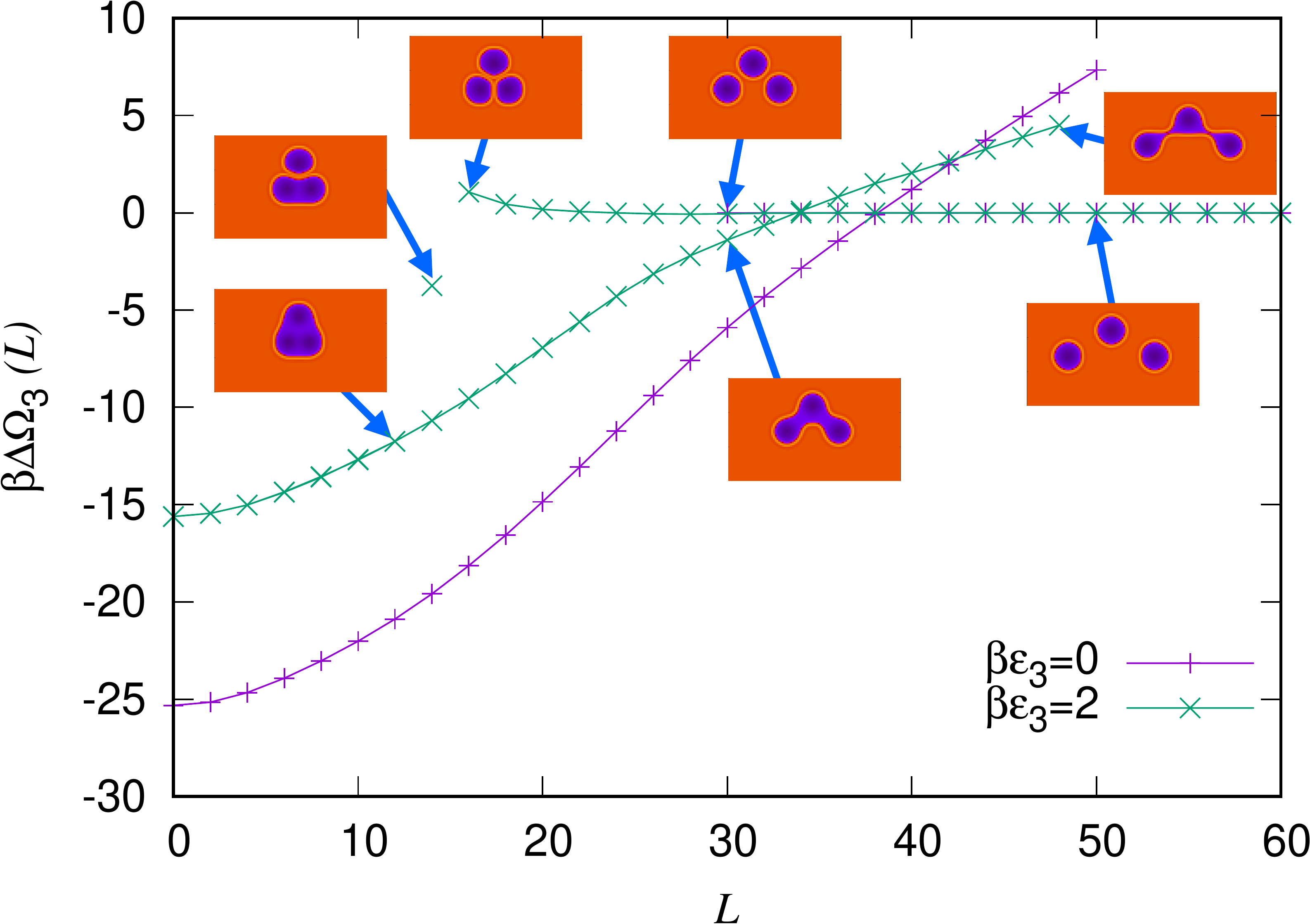}
    
    \caption{The three-body interaction potential $\Delta\Omega_3$, for droplet configurations where the upper droplet position is fixed, while the distance $L$ between the lower pair of droplets is varied, for $\beta\epsilon_3=0$ (purple) and $\beta\epsilon_3=2$ (green). For the $\beta\epsilon_3=2$ case, the inset plots display snapshots of the alcohol density profile for various $L$ on the three different solution branches. The centre of the upper drop is a distance $16\sigma$ above the mid-point of the line connecting the centres of the lower pair. The total number of oil molecules in the system $N_{\rm o}=1200$, while the chemical potentials of the water and alcohol are $\beta\mu_{\rm w}=-3.5$ and $\beta\mu_{\rm a}=-4$, respectively (i.e.\ bulk at point C in Fig.~\ref{fig:phase_diag}).}
    \label{fig:three_body}
\end{figure}

The external potential in Eq.~\eqref{eq:ext_pot_o_2D} used to determine the two-body interaction $\Delta\Omega_2(L)$ fixes the centres of the pair of oil droplets at the locations $\ii_1=\ii_c+(\frac12L,0)$ and $\ii_2=\ii_c+(-\frac12L,0)$, where $\ii_c=(40,40)$ corresponds to the lattice site of the centre of the (square) simulation box.
Recall that $\xx_i=\ii_i\sigma$, where $\sigma$ is the lattice spacing.
To determine the three-body interaction potential between a triplet of droplets, we add to the potential in Eq.~\eqref{eq:ext_pot_o_2D} an additional Gaussian well centred at $\ii_3$.
For simplicity, we keep the centres of two of the droplets at points $\ii_1$ and $\ii_2$ (the same as in our calculations for pairs of droplets) and locate the third droplet a distance of $15\sigma$ above the mid-point of the line between the first two droplets, i.e.\ with centre at $\ii_3=\ii_c+(0,15)$.
In Fig.~\ref{fig:three_body} we display the three-body potential $\Delta\Omega_3$, defined in Eq.~\eqref{eq:3body}, for varying $L$, i.e.\ for varying distance between the lower pair of droplets, keeping the upper one fixed.
The chemical potentials of the water and alcohol are $\beta\mu_{\rm w}=-3.5$ and $\beta\mu_{\rm a}=-4$, the same as for the cases considered in Fig.~\ref{fig:drop_pots_surf}, while the total number of oil molecules in the system $N_{\rm o}=1200$.
With this value for $N_{\rm o}$, when well-separated, the three droplets that form are very similar in size to the corresponding pairs of droplets considered in Fig.~\ref{fig:drop_pots_surf}.

Comparing the three-body interaction potential $\Delta\Omega_3(L)$ displayed in Fig.~\ref{fig:three_body} for the surfactant model with $\beta\epsilon_3=2$ and the `regular' oil-water-alcohol system with $\epsilon_3=0$, we observe that the overall energy scale $\Delta\Omega_3(L=0)$ is larger for the $\epsilon_3=0$ case.
This is because the interfacial tension $\gamma_{\rm ow}$ is larger for $\epsilon_3=0$ than it is for $\epsilon_3>0$, and it is the value of $\gamma_{\rm ow}$ that sets the overall energy scale.
For both cases, there is a jump in the potential $\Delta\Omega_3(L)$ corresponding to configurations where the droplets are joined together or not, with an associated hysteresis interval, just like for the two-body potentials displayed in Fig.~\ref{fig:drop_pots_surf}.

Like in Fig.~\ref{fig:drop_pots_surf}, we observe in Fig.~\ref{fig:three_body} that when $\beta\epsilon_3=2$ there is a free-energetic barrier to surmount in order for the droplets to merge, i.e.\ as droplets approach one another, the effective interaction potential is repulsive.
We also observe that in this case there are three branches to the free energy for this set of configurations, corresponding to (i) all droplets separate (ii) a pair of the droplets bridged and (iii) all three droplets bridged to one another.
We have not considered all possible configurations of the three droplets, but the results here show much of what is possible.
It is straightforward to calculate the value of $\Delta\Omega_3(\ii_1,\ii_2,\ii_3)$, defined in Eq.~\eqref{eq:3body}, for any configuration $(\ii_1,\ii_2,\ii_3)$ of the three droplets.
The advantage of our lattice DFT is that the calculations are relatively quick to perform on modern computers and so one could build on our approach to investigate the dynamics of droplets based on these potentials, by moving the droplets around and calculating the new potential at each timestep `on the fly', somewhat akin to what is done in the Car-Parrinello simulation method \cite{car1985unified}.
Here, we take a different approach to consider droplet dynamics, described below in Sec.~\ref{sec:dynamics}.

Comparing the results in Figs.~\ref{fig:drop_pots_surf} and~\ref{fig:three_body}, we can also infer that triplet droplet interactions cannot be expressed as a sum of pairs of two-body interactions.
This can be seen from considering the value of the two-body and three-body potentials when $L=0$.
For a pair of droplets with $\epsilon_3=0$, we can read off from Fig.~\ref{fig:drop_pots_surf} that $\beta\Delta\Omega_2(L=0)\approx-14$.
For three droplets, assuming they interact as three pairs, this would give a value for the three-body interaction at $L=0$ for all pairs of $-14\times 3 = -42$.
However, Fig.~\ref{fig:three_body} shows that this estimate is very wrong, where we see that in fact $\beta\Delta\Omega_2(L=0)\approx-25$. Clearly, droplet interactions are not pairwise-additive.

\section{Coarsening and dynamics of droplet coalescence}
\label{sec:dynamics}

\begin{figure}
    \centering
    \includegraphics[width=0.99\textwidth]{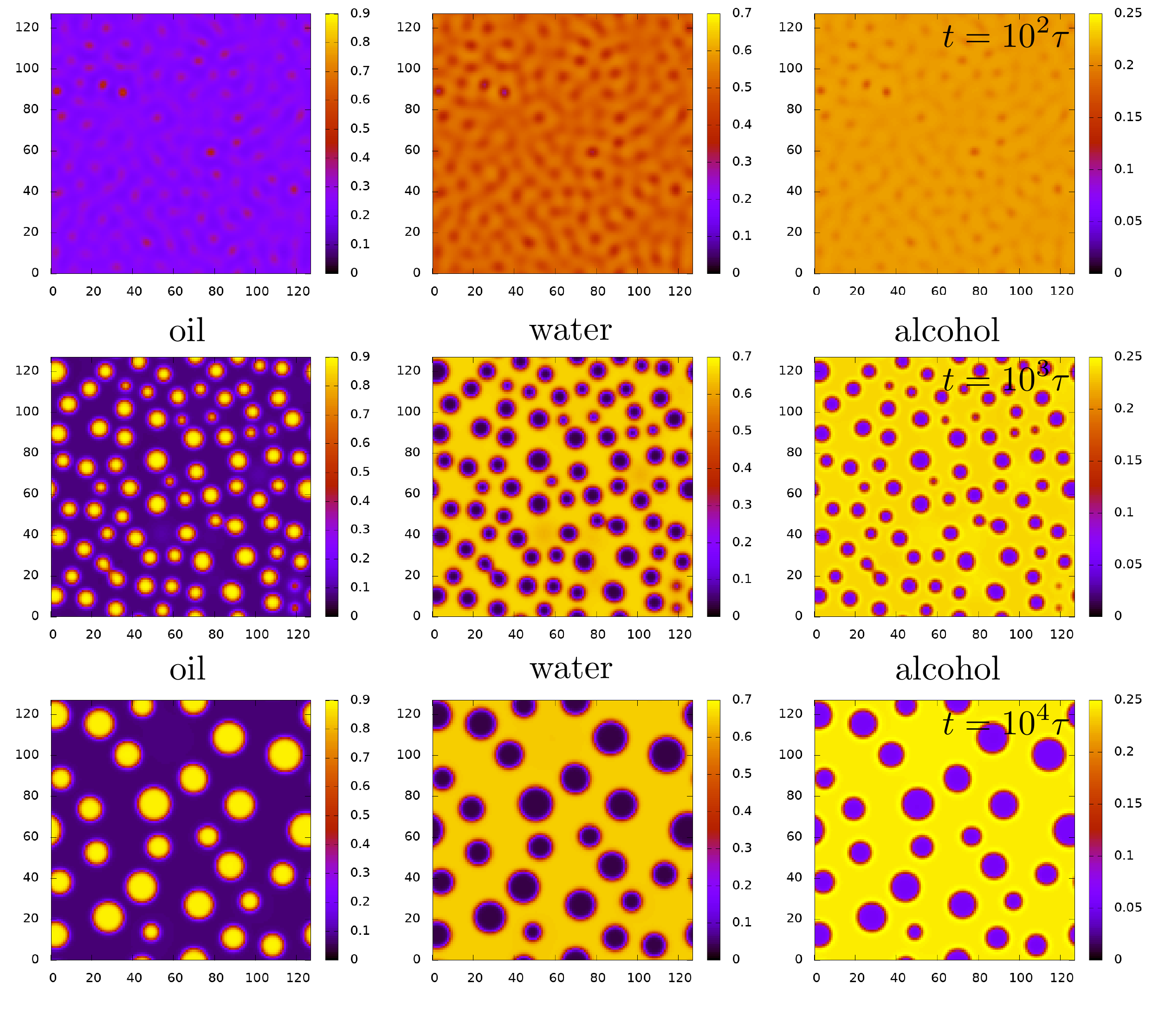}
    
    \caption{Time evolution of the density profiles after a quench, with the oil profiles on the left, the water in the middle and alcohol on the right, for the case when $\epsilon_3=0$. The time $t=0$ state consists of uniform densities $n_{\rm a}=0.21$, $n_{\rm o}=0.25$ and $n_{\rm w}=0.5$, with a small amplitude random noise field added. The profiles displayed are for $t=10^2\tau$, $t=10^3\tau$ and $t=10^4\tau$, where $\tau$ is the Brownian timescale. The final $t\to\infty$ equilibrium state (not displayed) corresponds to bulk two-phase coexistence.}
    \label{fig:spindec_e3_0}
\end{figure}

\begin{figure}
    \centering
    \includegraphics[width=0.99\textwidth]{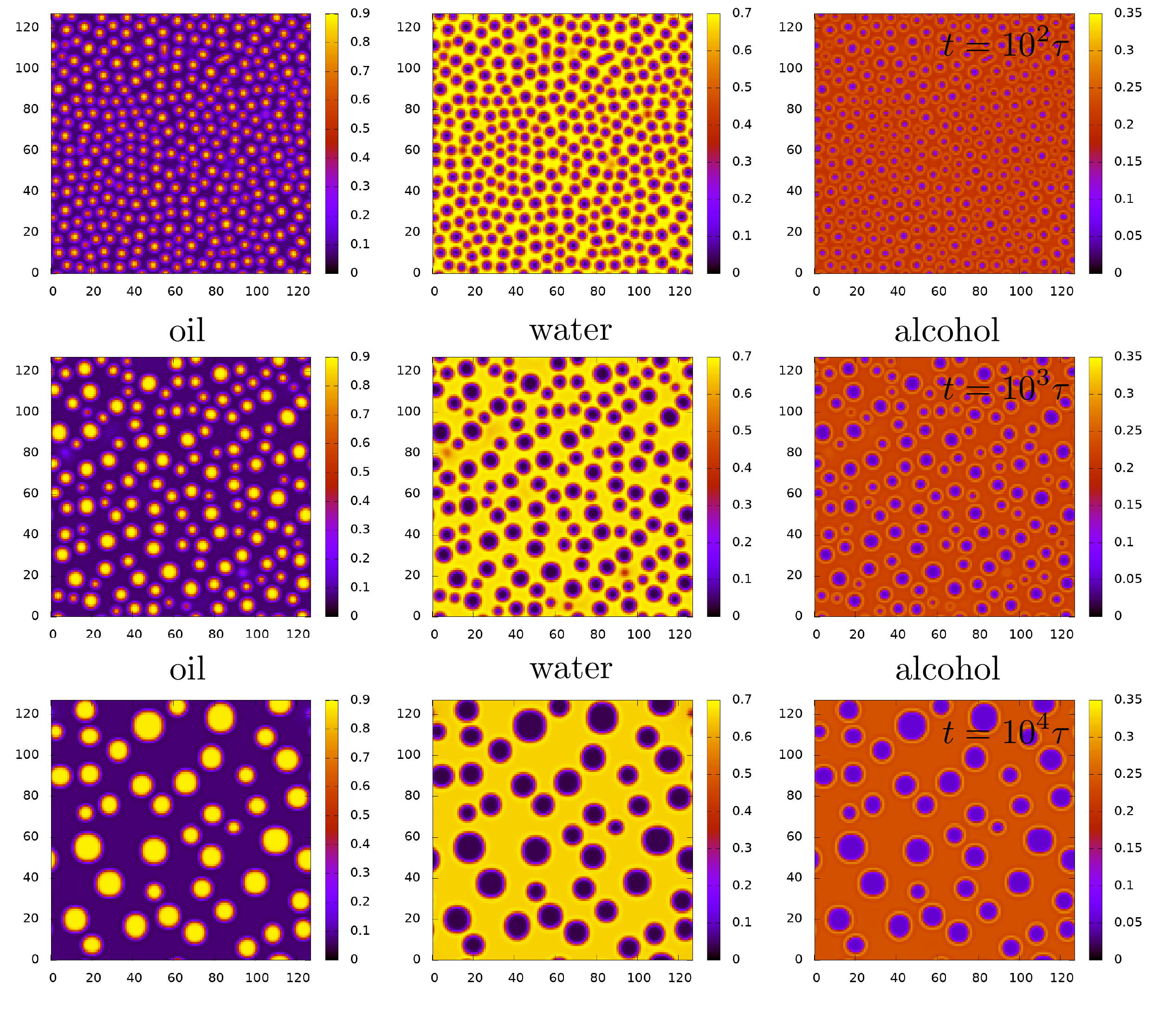}
    
    \caption{This is the same as Fig.~\ref{fig:spindec_e3_0}, except here $\beta\epsilon_3=2$.}
    \label{fig:spindec_e3_2}
\end{figure}

\begin{figure}
    \centering
    \includegraphics[width=0.49\textwidth]{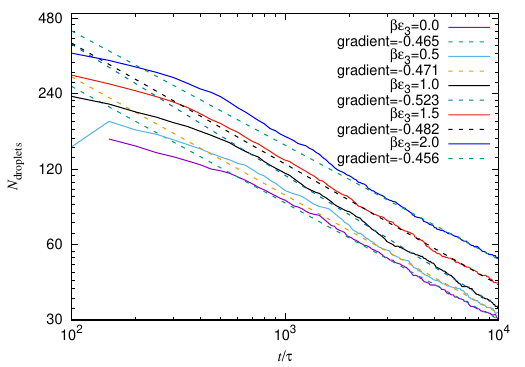}
    \includegraphics[width=0.49\textwidth]{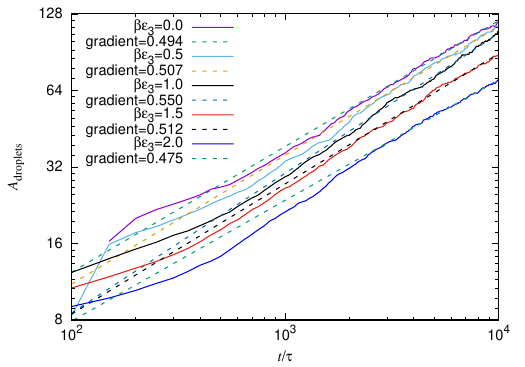}
    
    \caption{Plots of the average number of droplets and average droplet area corresponding to the results in Figs.~\ref{fig:spindec_e3_0} and \ref{fig:spindec_e3_2} and a few intermediate cases. Note that these are calculated by averaging over 5 different independent runs with different realizations of the initial random noise field in each case.}
    \label{fig:quench_plot}
\end{figure}

To investigate the fluid dynamics, we assume that the density profiles $n_p$ are now functions of time $t$ and that the time evolution of these three coupled density fields can be obtained from DDFT \cite{marconi1999dynamic, archer2004dynamical, espanol2009derivation, te2020classical}, with governing equations
\begin{equation}
\frac{\partial n_p}{\partial t}=\nabla\cdot\left[M_p n_p\nabla\frac{\delta F}{\delta n_p}\right],
    \label{eq:DDFT}
\end{equation}
where $F$ is the free energy in Eq.~\eqref{eq:helmholtz_continuum} and where the mobility coefficients $M_p=\beta D_p$, where $D_p$ are the diffusion coefficients for molecular species $p$.
With the approximation onto the lattice discussed above in Sec.~\ref{sec:3} for the free energy $F$, the $\nabla$ operators in Eq.~\eqref{eq:DDFT} represent finite difference approximations.
We use here the Euler algorithm based finite difference scheme developed in Ref.~\onlinecite{chalmers2017dynamical} to obtain the time evolution of the density profiles. To further simplify, we also assume that the diffusion coefficients for all three species are equal, $D_p=D$ for all $p$, so that the timescale governing the time-evolution of our system is the Brownian timescale $\tau=\sigma^2/D$.
Equation \eqref{eq:DDFT} assumes that the dynamics is isothermal and that inertial effects are negligible.
To go beyond this, one could use the DDFT of Refs.~\onlinecite{archer2006dynamical, archer2009dynamical, goddard2012general, goddard2012unification, stierle2021hydrodynamic}.
However, for present purposes, where we are largely interested in observing the influence of the alcohol/surfactant on the stability of droplets, overdamped DDFT \eqref{eq:DDFT} is sufficient.

In Figs.~\ref{fig:spindec_e3_0} and \ref{fig:spindec_e3_2} we present DDFT results corresponding to a quench to the state point with average densities $n_{\rm a}=0.21$, $n_{\rm o}=0.25$ and $n_{\rm w}=0.5$.
This is a state point inside the spinodal, that is much closer to the water-rich binodal than to the coexisting oil-rich state point -- see Fig.~\ref{fig:phase_diag}.
Due to this, as the mixture phase separates, it quickly forms droplets of oil in a background of the majority water-rich phase (rather than a bicontinuous network-like structure, which is what we observe for a quench to the regions around the mid-point of the coexistence tie-lines between the binodals).
The $t=0$ initial condition for our DDFT computations sets the densities equal to the average values plus a small amplitude random noise at each lattice site.
At this state point the system is linearly unstable and so some of the small amplitude perturbations grow in amplitude over time, leading to phase separation via spinodal decomposition \cite{bray2002theory, onuki2002phase}.
Figure~\ref{fig:spindec_e3_0} shows results for the original ouzo model (i.e.\ with $\beta\epsilon_3=0$), while Fig.~\ref{fig:spindec_e3_2} is for the surfactant model with $\beta\epsilon_3=2$, while all other parameters are the same.
In both Figs.~\ref{fig:spindec_e3_0} and \ref{fig:spindec_e3_2} we observe that shortly after the quench, the phase separation leads to the formation of numerous small oil droplets, surrounded by the majority water-rich phase.
Over time some of the droplets merge,
illustrating clearly what the results of Sec.~\ref{sec:results} lead us to expect, i.e.\ that the droplets have an effective attraction to each other.
In parallel with the merging events, we also observe some coarsening via Ostwald ripening.
Visual comparison of Figs.~\ref{fig:spindec_e3_0} and \ref{fig:spindec_e3_2} also shows that the surfactant makes the droplets more stable over time (i.e.\ remaining smaller and more numerous), again in agreement with what one would expect based on the effective interaction potentials calculated in Sec.~\ref{subsec:surfactant_pots}.

To quantify the above observations, we have developed a droplet analysis algorithm for counting the number of droplets over time and for calculating the area of each of the droplets (recall we are treating the system as 2D, so the area is a measure of droplet size).
The number and areas of the oil droplets are determined using Matlab's \texttt{regionprops} command \cite{regionprops}.
As a first step, a threshold is applied to the oil density profiles, with lattice values greater than 0.5 set to 1, and values lower than 0.5 set to 0.
Note that the results are not particularly sensitive to this threshold value.
The number of connected regions (i.e., droplets) is then determined by connecting all lattice sites with value 1 to any of their nearest and next-nearest lattice sites that also have the value 1.
The area of each droplet is then defined as the number of cells that each connected region contains.
Note that this method leads to some (small) artefacts due to the periodic boundary conditions, but this only becomes significant when the number of droplets is very small.

Results from this analysis are displayed in Fig.~\ref{fig:quench_plot}, where we present results for the average number of droplets $N_\mathrm{droplets}$ and average area $A_\mathrm{droplets}$ over time since the quench (at time $t=0$), for $\beta\epsilon_3=0$, 0.5, 1, 1.5 and 2.
In each case, these are calculated by averaging over five independent runs, each with a different realization of the initial random noise field.
The plot of $N_\mathrm{droplets}$ over time shows that as the value of $\epsilon_3$ is increased, the average number of droplets at any given time after the quench, is increased.
Similarly, the plot of the average area $A_\mathrm{droplets}$ shows that these drops are correspondingly smaller.
These plots have a logarithmic time-axis, which allow us to observe (albeit over only two decades in time) that the change over time has a power-law behaviour, with an exponent (given in the key) that depends weakly on the value of $\epsilon_3$.
Such variations of the exponent are to be expected \cite{konig2021two}, since we are away from the critical composition.

\section{Concluding remarks}
\label{sec:conc}

In this paper we have developed a general widely-applicable DFT-based method for calculating the effective interaction potential (or `potential of mean force') between liquid droplets in immiscible liquid mixtures.
We use a small external potential to constrain the centres of the droplets to the specified distances apart.
The method can be used to determine the pair interaction potential $\Delta\Omega_2(L)$ and also its multi-drop generalisation $\Delta\Omega_m(\mathbf{x}_1,\cdots,\mathbf{x}_m)$, for the case of $m$ different droplets.
Here, we have applied our method to oil-water mixtures and also to ternary alcohol-oil-water mixtures.
The alcohol behaves as a weak surfactant and we find that its presence decreases the overall strength of the effective interaction potential between pairs of oil droplets, because it decreases the oil-water interfacial tension.
We have also been able to vary the surfactant-like properties of the alcohol.
Increasing the affinity of the `alcohol' towards the oil-water interface, making its behaviour in our model akin to that of a stronger surfactant, leads to the effective interaction potential between the oil droplets becoming repulsive, with a free-energy barrier to overcome, for droplets to coalesce.
The barrier height is only a few $k_BT$ for the small oil droplets considered here.
This would easily be overcome by thermal fluctuations.
However, since the size of the barrier scales with droplet size, this illustrates how surfactants can stabilize oil droplets in water.

As remarked in the Introduction, the effective potentials $\Delta\Omega_2(L)$ that we calculate have two different branches, corresponding to a jump in $f_2=-\partial\Delta\Omega_2(L)/\partial L$, giving a jump in the force as droplets merge.
This is qualitatively very similar to the results obtained in the AFM experiments \cite{israelachvili2011intermolecular, gunning2004atomic, sun2021unraveling}.
Likewise, molecular dynamics simulation results for the potential of mean force are qualitatively very similar to those observed here for our surfactant model with $\epsilon_3>0$ \cite{sun2021unraveling}. See also Refs.~\onlinecite{arbabi2025collision, pak2018free, inada2025molecular}.

One aspect that we have entirely neglected here is that of the hydrodynamics of two droplets approaching one another in a surrounding fluid.
Relevant for larger drops, Refs.~\onlinecite{chan2011film, berry2017mapping} discus some of the subtleties of how the fluid flow between a pair of droplets affects the force between them as they approach each other.
Ref.~\onlinecite{zhang2023comprehensive} also gives a broad discussion of the behaviour of oil droplets in water.
For a general review of how surfactant-like species can stabilize oil droplets in water, with a discussion of the interactions, including influence of charges, see Ref.~\onlinecite{dickinson2010flocculation}.

In the present study, we used a simple lattice-DFT that gives a good account of the bulk and interfacial thermodynamics \cite{archer2024experimental, sibley2025coexisting}. However, lattice-DFT can be improved and made more accurate by borrowing ideas from continuum DFT (e.g.\ fundamental measure theory \cite{hansen2013theory, roth2010fundamental}) in order to improve how the lattice-DFT describes the excluded volume correlations \cite{maeritz2021droplet, maeritz2021density, zimmermann2024lattice}.
If one requires a more accurate and detailed description of how the molecular correlations affect the structure of droplets (see e.g.\ the recent study \cite{gul2025classical}), then ultimately one should move off-lattice and implement the general method presented here together with a suitable continuum DFT.

\section*{Acknowledgements}

Both B.~D.~Goddard and A.~J.~Archer gratefully acknowledge the support for parts of this work from the London Mathematical Society and the Loughborough University Institute of Advanced Studies. The authors would like to thank Tapio Ala-Nissila, David Fairhurst, Marco Mazza and Fouzia Ouali for valuable discussions.

\section*{Appendix}

We show here how to obtain the mapping between the lattice description and the continuum description and give more details of the derivation of Eq.~\eqref{eq:lattice_to_continuum}.
The gist of the argument is best seen by considering first the one dimensional (1D) version of the lattice DFT.

In the 1D version of the model, the lattice index $\ii\equiv i$, with the corresponding pair interaction matrix
\begin{equation}
\label{eq:e_ij_1D}
c_{{ij}}  = 
  \begin{cases} 
   1 & \text{if }j\in {NN i}, \\
        0   & \text{otherwise}.
         \end{cases}
\end{equation}
In other words, $c_{ij}=1$ when $j=i\pm1$ and $c_{ij}=0$, otherwise.
Thus, the pair interaction term in the free energy can be written as
\begin{eqnarray}
  - \sum_{\ii,\j}\e_{\ii\j}^{pq} n_\ii^p n_\j^q  &=& - \epsilon_{pq} \sum_{i}\sum_j c_{ij} n_i^p n_j^q\nonumber \\
  &=& - \epsilon_{pq} \sum_{i} n_i^p (n_{i+1}^q+n_{i-1}^q)\nonumber \\
  &=& - \epsilon_{pq} \sum_{i} n_i^p (n_{i+1}^q-2n_i^q+n_{i-1}^q)-\epsilon_{pq} \sum_{i} n_i^p 2 n_i^q\nonumber \\
  &=& - \epsilon_{pq} \sigma^2 \sum_{i} n_i^p \frac{n_{i+1}^q-2n_i^q+n_{i-1}^q}{\sigma^2}-\epsilon_{pq} \sum_{i} n_i^p 2 n_i^q\nonumber \\
  &\approx& - \epsilon_{pq} \sigma^2 \sum_{i} n_i^p \frac{d^2 n_{i}^q}{dx^2}-2\epsilon_{pq} \sum_{i} n_i^p n_i^q
  \label{eq:1D}
\end{eqnarray}
where $\sigma=1$ is the lattice spacing and we have used the finite difference approximation for the second derivative $\frac{d^2f}{dx^2}\approx\frac{f(x+h)-2f(x)+f(x-h)}{h^2}$ with $x=i\sigma$ and $h=\sigma$ to obtain the last line in Eq.~\eqref{eq:1D}. Notice also that the factor 2 in the last term is obtained from $\sum_jc_{ij}=2$.

If we now map the above discrete system onto the continuum, replacing $\sigma=1$, $\sum_i\to\int dx$ and $n^p_i\to n_p(x)$, we obtain:
\begin{eqnarray}
  - \sum_{\ii,\j}\e_{\ii\j}^{pq} n_\ii^p n_\j^q  &\to& \int \left[ -\epsilon_{pq} n_p(x) \frac{d^2 n_q(x)}{dx^2}-2\epsilon_{pq} n_p(x) n_q(x)\right]dx\nonumber \\
  &=& \int \left[ \epsilon_{pq} \frac{d n_p(x)}{dx} \frac{d n_q(x)}{dx}-2\epsilon_{pq} n_p(x) n_q(x)\right]dx,
  \label{eq:1D_continuum}
\end{eqnarray}
where to obtain the second line we integrate by parts assuming either periodic boundary conditions or that $n_p=0$ on the boundaries.

Returning to the 3D model, the generalization of the above argument proceeds as follows. The pair interaction term in the free energy can be written as
\begin{eqnarray}
  - \sum_{\ii,\j}\e_{\ii\j}^{pq} n_\ii^p n_\j^q  &=& - \epsilon_{pq} \sum_{\ii}\sum_\j c_{\ii\j} n_\ii^p n_\j^q\nonumber \\
  &=& - \epsilon_{pq} \sum_{\ii} n_\ii^p \left(\sum_{NN\ii}n_{\j}^q+\frac{3}{10}\sum_{NNN\ii}n_{\j}^q+\frac{1}{20}\sum_{NNNN\ii}n_{\j}^q\right)\nonumber \\
  &=& - \epsilon_{pq}\frac{48\sigma^2}{20} \sum_{\ii} n_\ii^p \frac{1}{48\sigma^2}\left(20\sum_{NN\ii}n_{\j}^q+6\sum_{NNN\ii}n_{\j}^q+\sum_{NNNN\ii}n_{\j}^q-200n_\j^q\right)\nonumber \\ &\,&\hspace{5cm}-10\epsilon_{pq}\sum_\ii n_\ii^p n_\j^q\nonumber \\
  &=& - \epsilon_{pq}\frac{12\sigma^2}{5} \sum_{\ii} n_\ii^p \nabla^2n_\ii^p
  -10\epsilon_{pq}\sum_\ii n_\ii^p n_\j^q
  ,
  \label{eq:aa}
\end{eqnarray}
where $\sum_{NN\ii}$ denotes the sum over the 6 nearest neighbour lattice sites of site $\ii$, $\sum_{NNN\ii}$ denotes the sum over the 12 next-nearest neighbours of $\ii$ and $\sum_{NNNN\ii}$ denotes the sum over the 8 next-next-nearest neighbours of $\ii$.
Note that we used Eq.~\eqref{eq:c_ij} to obtain the second line in Eq.~\eqref{eq:aa}.
To obtain the third line, we used the isotropic discretisation of the Laplacian derived in Ref.~\onlinecite{kumar2004isotropic}:
\begin{equation}
    \nabla^2n_\ii^q = \frac{1}{48h^2}\left( 
    20\sum_{NN\ii}n_{\j}^q+6\sum_{NNN\ii}n_{\j}^q+\sum_{NNNN\ii}n_{\j}^q - 200n_{\j}^q
    \right),
\end{equation}
where $h$ is the grid spacing, which here we set $h=\sigma=1$. The last line of Eq.~\eqref{eq:aa} is the result given in Eq.~\eqref{eq:lattice_to_continuum} of the main text.

Mapping the above discrete system onto the continuum [c.f.~Eq.~\eqref{eq:1D_continuum}], replacing $\ii\sigma\to\xx$, $\sigma=1$, $\sum_\ii\to\int d\xx$ and $n^p_\ii\to n_p(\xx)$, we obtain:
\begin{eqnarray}
  - \sum_{\ii,\j}\e_{\ii\j}^{pq} n_\ii^p n_\j^q  &\to& \int \left[ -\frac{12}{5}\epsilon_{pq} n_p(\xx) \nabla^2 n_q(\xx)-10\epsilon_{pq} n_p(\xx) n_q(\xx)\right]d\xx\nonumber \\
  &=& \int \left[ \frac{12}{5}\epsilon_{pq} \nabla n_p(\xx)\cdot \nabla n_q(\xx)-10\epsilon_{pq} n_p(\xx) n_q(\xx)\right]d\xx,
  \label{eq:3D_continuum}
\end{eqnarray}
where again, to obtain the second line, we integrate by parts assuming either periodic boundary conditions or that $n_p=0$ or that $\nabla n_p(\xx) \cdot \boldsymbol{\nu}(\xx)=0$ on the boundaries, where $\boldsymbol{\nu}(\xx)$ is the normal vector on the boundary.


%

\end{document}